\documentclass[%
 reprint,
superscriptaddress,
aps, pra, twocolumn,
amsmath,amssymb,
]{revtex4-2}

\usepackage{amsmath}
\usepackage{braket}
\usepackage{xcolor}
\usepackage{graphicx}
\usepackage{hyperref}
\usepackage{ulem}

\newcommand{\hc}{\mathrm{h.c.}}
\newcommand{\cc}{\mathrm{c.c.}}
\newcommand{\omax}{\Omega_\mathrm{max}}
\newcommand{\otr}{\omega_\mathrm{tr}}

\renewcommand{\Re}{\mathrm{Re}}

\newcommand{\ton}{\tau_{\mathrm{on}}}

\newcommand{\edit}[2]{#2}
\renewcommand{\emph}[1]{{\it{#1}}}

\begin{document}


\title{Optimizing Rydberg Gates for Logical Qubit Performance}



\author{Sven Jandura}
\affiliation{University of Strasbourg and CNRS, CESQ and ISIS (UMR 7006), aQCess, 67000 Strasbourg, France}

\author{Jeff D. Thompson}
\affiliation{Department of Electrical and Computer Engineering, Princeton University, Princeton, NJ 08544, USA}

\author{Guido Pupillo}
\affiliation{University of Strasbourg and CNRS, CESQ and ISIS (UMR 7006), aQCess, 67000 Strasbourg, France}



\date{\today}

\begin{abstract}
Robust gate sequences are widely used to reduce the sensitivity of
gate operations to experimental imperfections. Typically, the
optimization minimizes the average gate error, however, recent work in
quantum error correction has demonstrated that the performance of
encoded logical qubits is sensitive to not only the average error
rate, but also the type of errors that occur. Here, we present a
family of Rydberg blockade gates for neutral atom qubits that are
robust against two common, major imperfections: intensity
inhomogeneity and Doppler shifts.
These gates outperform existing gates for moderate or large
imperfections. We also consider the logical performance of
these gates in the context of an erasure-biased qubit based on
metastable $^{171}$Yb. In this case, we observe that the robust gates
outperform existing gates for even very small values of the
imperfections, because they maintain the native large bias towards
erasure errors for these qubits. These results significantly reduce the laser stability and atomic temperature requirements to achieve fault-tolerant quantum computing with neutral atoms. The approach of optimizing gates for logical qubit performance may be applied to other qubit platforms.
\end{abstract}


\maketitle

\section{Introduction}
\label{sec:introduction}
Neutral atoms have emerged as a leading platform for digital quantum simulations and quantum computing \cite{bloch_many-body_2008, saffman_quantum_2010, browaeys_many-body_2020, henriet_quantum_2020, morgado_quantum_2021}. Optical tweezer arrays enable hundreds of atoms to be configured in arbitrary, reconfigurable geometries \cite{endres_atom-by-atom_2016, barredo_atom-by-atom_2016,ohl_de_mello_defect-free_2019, barredo_synthetic_2018, schlosser_large-scale_2019}. Laser excitation to Rydberg states allows strong, controllable interactions via the Rydberg blockade \cite{jaksch_fast_2000}, which enables studies of many-body dynamics \cite{browaeys_many-body_2020} or quantum logic operations \cite{isenhower_multibit_2011, wilk_entanglement_2010, levine_parallel_2019, graham_rydberg-mediated_2019, madjarov_high-fidelity_2020, ma_universal_2022, schine_long-lived_2022}. Recent work has demonstrated the execution of programmable quantum circuits on multi-qubit systems \cite{bluvstein_quantum_2022, graham_multi-qubit_2022}. 

One of the most important directions for continued development of neutral atom processors is improving the fidelity of two-qubit Rydberg blockade gates \cite{saffman_quantum_2016}. The best demonstrated Bell state fidelity using hyperfine qubits is around $F = 0.98$ \cite{levine_parallel_2019,fu_high-fidelity_2022}, and Bell state fidelities of $F = 0.991$ have been demonstrated between ground and Rydberg states \cite{madjarov_high-fidelity_2020}. While the achievable gate fidelity is fundamentally limited by the Rydberg state lifetime and the achievable laser intensity, currently demonstrated fidelities are dominated by technical imperfections including Doppler shifts from finite atomic temperature, spatial inhomogeneity of the laser intensity and frequency and intensity noise inherent to the laser \cite{saffman_quantum_2016,de_leseleuc_analysis_2018,levine_high-fidelity_2018, graham_rydberg-mediated_2019}.

With the recent demonstration of rudimentary fault-tolerant quantum operations in several qubit platforms \cite{egan_fault-tolerant_2021, ryan-anderson_realization_2021, postler_demonstration_2022, abobeih_fault-tolerant_2022, krinner_realizing_2022, zhao_realization_2022}, there is a significant interest in predicting and optimizing the performance of not only physical qubit operations, but also logical qubit operations. The performance of quantum error correction depends strongly on the types of errors that occur. For example, qubits strongly biased towards certain Pauli errors \cite{aliferis_fault-tolerant_2008,lescanne_exponential_2020,grimm_stabilization_2020, darmawan_practical_2021,cong_hardware-efficient_2022}, or erasure errors \cite{stace_thresholds_2009, wu_erasure_2022, kubica_erasure_2022} can have significantly higher threshold error rates than unbiased qubits. However, maintaining this improved performance in a realistic architecture requires that the noise structure is preserved in a quantum circuit including realistic physical gate operations and the influence of imperfections \cite{puri_bias-preserving_2020,cong_hardware-efficient_2022}.

In this work, we present a family of two-qubit Rydberg blockade gates that offer significantly improved performance in the presence of experimental imperfections. These gates are derived by combining analytic reasoning and quantum optimal control techniques \cite{jandura_time-optimal_2022, pagano_error_2022} to produce experimentally realizable pulse shapes that are robust against laser amplitude inhomogeneity, Doppler shifts from finite atomic temperature, or both \cite{mitra_robust_2020, goerz_robustness_2014}. These pulses do not require individual addressability of the atoms and can be implemented using only a smooth modulation of the laser phase in time. In comparison to existing gates, these pulses reduce the error from intensity inhomogeneity and Doppler shifts by more than two orders of magnitude, at the expense of longer gate duration. Including the finite lifetime of the Rydberg state, we find that the robust pulses significantly outperform existing gates over a very wide parameter range.

We also evaluate the performance of the robust gates in the context of a logical qubit. For this purpose, we adopt the XZZX surface code \cite{bonilla_ataides_xzzx_2021} and the strongly erasure-biased metastable  $^{171}$Yb qubit \cite{wu_erasure_2022}. Surprisingly, we identify a large and experimentally relevant range of imperfections where the robust gates \emph{decrease} the average gate fidelity at the physical qubit level (compared to existing gates), but \emph{improve} the logical qubit performance by many orders of magnitude. This arises because the robust gates constrain imperfections to only cause Rydberg leakage, and not errors in the computational space. Since Rydberg leakage can be converted into erasure errors, these gates allow the predicted high erasure bias and threshold error rate of the metastable $^{171}$Yb qubit to be extended to errors from laser amplitude fluctuations and Doppler shifts.

There are two main implications of this work. First, we show that robust Rydberg blockade gates give rise to significantly improved physical and logical operation fidelity in the presence of technical imperfections, which significantly enhances the prospects for fault tolerant quantum computing (FTQC) with neutral atoms. Second, we demonstrate that physical and logical error rates can \edit{diverge}{differ significantly} in the context of qubits with biased noise, and that optimizing gates specifically for logical-level optimization can yield dramatic improvements. This insight may be applied to other qubit platforms with biased noise \cite{aliferis_fault-tolerant_2008,lescanne_exponential_2020,grimm_stabilization_2020, darmawan_practical_2021,cong_hardware-efficient_2022}.

The remainder of this work is structured as follows: In Sec.~\ref{sec:level_scheme_and_hamiltonian} we introduce the Hamiltonian that we assume and the error sources that we consider. In Sec.\edit{~\ref{sec:amp_and_detuning_robust_pulses}}{~\ref{subsec:amp_robust_pulses}} we identify a gate which is robust against laser amplitude inhomogeneity, and prove \edit{}{in Sec.~\ref{subsec:detuning_robust_pulses}} that a similar gate robust against laser detuning \edit{}{generally} cannot exist. \edit{}{For two special cases of laser detuning we then show that robust gates nevertheless do exist: In Sec.~\ref{subsec:stark_shift_robust_pulses} we consider a laser detuning which arises due to an AC Stark shift and is thus correlated with the laser amplitude inhomogeniety, and identify robust pulses for this case.} In Sec.~\ref{sec:doppler_robust_pulses}, we propose a \edit{work-around}{detuning robust gate} based on reversing the sign of the detuning halfway through the gate, and observe that this can be realized for detunings arising from Doppler shifts by reversing the laser direction or exploiting the harmonic motion of trapped atoms. We then derive a gate that is robust to both Doppler and amplitude imperfections. In Sec.~\ref{sec:infidelities_in_a_realistic_error_model} we calculate the gate fidelities for the robust pulses using realistic experimental parameters, and in Sec.~\ref{sec:conditional_infidelity_and:logical_error_rate} we calculate the logical error rate for our robust pulses in a small surface code, in the context of erasure-biased metastable $^{171}$Yb qubits \cite{wu_erasure_2022}. We use insights from Sec.~\ref{sec:conditional_infidelity_and:logical_error_rate} in Sec.~\ref{sec:conditionally_robust_pulses} to design shorter pulses that are optimized for erasure bias but not average gate fidelity, which further improves the logical error rate in certain parameter regimes.

\section{Level Scheme and Hamiltonian}
\label{sec:level_scheme_and_hamiltonian}
\begin{figure}
\includegraphics[width=\linewidth]{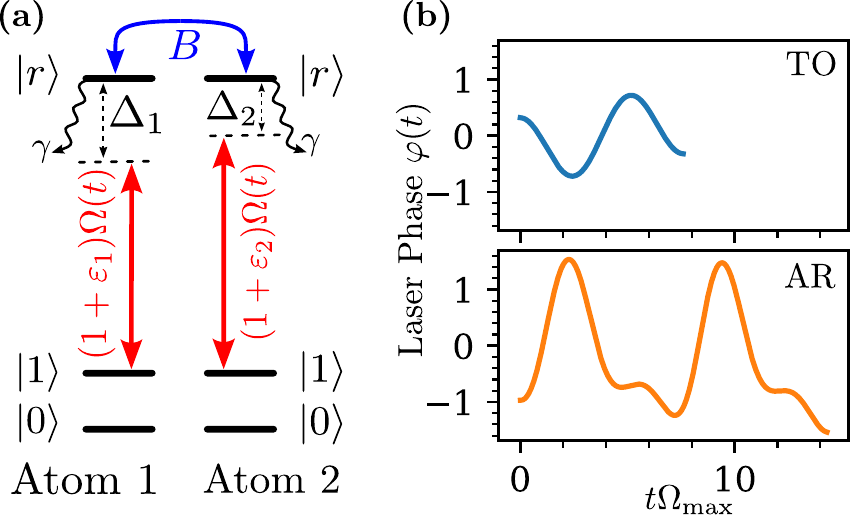}
\caption{\label{fig:level_scheme_and_pulses}a) Qubits are stored in computational basis states $\ket{0}$ and $\ket{1}$. The $\ket{1}$ state is coupled to a Rydberg state $\ket{r}$ with lifetime $1/\gamma$ by a laser with Rabi frequency $\Omega(t)$. The amplitude of the laser has an unknown relative deviation of $\varepsilon_1$ and $\varepsilon_2$ and an unknown detuning $\Delta_1$ and $\Delta_2$ for atom 1 and 2, respectively. The van der Waals interaction leads to an energy shift $B\gg|\Omega|$ if both atoms are in $\ket{r}$, preventing the simultaneous excitation of both atoms. b) The laser phase as a function of time for the time-optimal (TO) and amplitude-robust (AR) pulse implementing the CZ gate. The amplitude of the pulses is always given by $|\Omega(t)|=\Omega_{\mathrm{max}}$.}
\end{figure}
We assume the level scheme typical for Rydberg gates \cite{jaksch_fast_2000, isenhower_multibit_2011, levine_parallel_2019, mitra_robust_2020, saffman_symmetric_2020,sun_controlled_2020, shen_construction_2019} shown in Fig.~\ref{fig:level_scheme_and_pulses}(a). Each atom 
is modeled as a three level system with long-lived computational basis states $\ket{0}$ and $\ket{1}$ and an auxiliary Rydberg state $\ket{r}$ with finite lifetime $1/\gamma$. The $\ket{1}$ and $\ket{r}$ states of each atom are coupled by a single global laser pulse with Rabi frequency $\Omega(t)=|\Omega(t)|\exp[i\varphi(t)]$ of amplitude $|\Omega(t)|$ and phase $\varphi (t)$.  We require that at all times $|\Omega(t)| \leq \omax$, where $\omax$ is the maximally achievable Rabi frequency. A van-der-Waals interaction between the two atoms shifts the energy of the state $\ket{rr}$ by $B\gg |\Omega(t)|$, in the Rydberg blockade limit, such that the state $\ket{rr}$ is never populated.

We parameterize the experimental imperfections that can impact the gate in two ways. First, intensity fluctuations (\emph{i.e.}, from spatial inhomogeneity or power drifts of the laser field) give rise to a uncertain Rabi frequency at atom $i$ (for $i\in\{1,2\}$) of $(1+\varepsilon_i)\Omega(t)$.
Second, laser frequency errors or Doppler shifts can result in a detuning $\Delta_i$ from the $\ket{1}\leftrightarrow\ket{r}$ transition. We consider $\varepsilon_i$ and $\Delta_i$ as unknown but constant over the duration of the gate. We note that the phase of $\Omega$ may vary from shot to shot by an uncertain offset, because it depends on the position of the atoms. However, in contrast to work on single qubit gates \cite{whitlock_robust_2022}, this uncertainty does not lead to an error for two-qubit Rydberg blockade gates, as the population of the Rydberg state vanishes at the end of the gate, so that the additional phase can be absorbed into the definition of $\ket{r}$.     

The total Hamiltonian reads 
$H = H_{10}\oplus H_{01}\oplus H_{11}$, with 
\begin{eqnarray}
H_{10} &=& \frac{(1+\varepsilon_1)\Omega(t)}{2}\ket{10}\bra{r0} + \hc +  \Delta_1\ket{r0}\bra{r0} \label{eq:H10}\\
H_{01} &=& \frac{(1+\varepsilon_2)\Omega(t)}{2}\ket{01}\bra{0r} + \hc +  \Delta_2\ket{0r}\bra{0r}\\
H_{11} &=& \frac{(1+\varepsilon_1)\Omega(t)}{2}\ket{11}\bra{r1} + \hc + \Delta_1 \ket{r1}\bra{r1} \label{eq:H11_complicated} \\ \nonumber
&+& \frac{(1+\varepsilon_2)\Omega(t)}{2}\ket{11}\bra{1r} + \hc + \Delta_2\ket{1r}\bra{1r}
\end{eqnarray}
\edit{}{Eq.~\eqref{eq:H11_complicated} can be simplified to 
\begin{eqnarray}
H_{11} &=& \frac{\sqrt{2}(1+\varepsilon_+)\Omega(t)}{2}\ket{11}\bra{W_+} +\hc \label{eq:H11} \\ \nonumber
&+& \Delta_- \ket{W_+}\bra{W_-}\nonumber +  \hc   \\ \nonumber 
&+& \Delta_+ \left(\ket{W_+}\bra{W_+}+\ket{W_-}\bra{W_-}\right).    
\end{eqnarray}
}
Here, $\ket{W_\pm} =  \left[(1+\varepsilon_1)\ket{r1} \pm (1+\varepsilon_2)\ket{1r}\right]/\beta$, with $\beta=\sqrt{(1+\varepsilon_1)^2 + (1+\varepsilon_2)^2}$, while $\Delta_\pm= (\Delta_1\pm\Delta_2)/2$ and $\varepsilon_+ = (\varepsilon_1+\varepsilon_2)/2$. \edit{In Eq.~\eqref{eq:H11} we neglected terms in second order of $\varepsilon$ and $\Delta$.}{Note that Eq.~\eqref{eq:H11_complicated} and Eq.~\eqref{eq:H11} only agree up to terms in second order in $\epsilon$ and $\Delta$. Since $\epsilon$ and $\Delta$ are typically small deviations, this suffices for the rest of our analysis.}

\section{Amplitude- and Detuning- Robust pulses}
\label{sec:amp_and_detuning_robust_pulses}
\subsection{Amplitude Robust pulses}
\label{subsec:amp_robust_pulses}
We start by finding a pulse $\Omega(t)$ which is robust against amplitude deviations $\varepsilon_i \neq 0$, while $\Delta_i=0$. We expand the Hamiltonians in $\varepsilon$ 
as  $H_{10} = H_{10}^{(0)} + \varepsilon_1 H_{10}^{(1)}$, $H_{01} = H_{01}^{(0)} + \varepsilon_2 H_{01}^{(1)}$ and $H_{11} = H_{11}^{(0)} + \varepsilon_+ H_{11}^{(1)}$ and the quantum state when starting in $\ket{01}$ as $\ket{\psi_{10}} = \ket{\psi_{10}^{(0)}} + \varepsilon_1\ket{\psi_{10}^{(1)}}+\mathcal{O}(\varepsilon_1^2)$, with analogous expansions for $\ket{\psi_{10}}$ and $\ket{\psi_{11}}$. A pulse $\Omega(t)$ of duration $\tau$ implements a CZ gate\edit{}{, up to single-qubit rotations,} in the deviation-free case if for all $q \in \{10,01,11\}$ it holds that $\ket{\psi_q^{(0)}(\tau)} = e^{i\theta_q}\ket{q}$ with $\theta_{11}-\theta_{10}-\theta_{01} = (2n+1)\pi$ for an integer $n$. We measure the fidelity of a gate via the Bell-state fidelity\edit{}{, a commonly used fidelity measure on the Rydberg platform \cite{levine_parallel_2019, graham_rydberg-mediated_2019, theis_high-fidelity_2016, robicheaux_photon-recoil_2021}.}
\begin{equation}
F = \frac{1}{16}\left|1+\sum_{q\in\{10,01,11\}} e^{-i\theta_q}\braket{q|\psi_q^{(0)}}\right|^2.
\label{eq:fidelity}
\end{equation}
\edit{}{The differences between the Bell-state fidelity and the average gate fidelity \cite{pedersen_fidelity_2007}, a different common fidelity measure, are discussed in Ref.~\cite{jandura_time-optimal_2022}.}

We say that the pulse is robust against amplitude deviations if $\ket{\psi_q^{(1)}}=0$ for all $q$, such that the leading term of the deviation of $\ket{\psi_q}$ from $e^{i\theta_q}\ket{q}$ is quadratic in $\varepsilon_i$. 
To find the robust pulse we minimize the cost function
\begin{equation}
J = 1-F + \sum_q \braket{\psi_q^{(1)}|\psi_q^{(1)}}
\label{eq:cost_function}
\end{equation}
using a quantum optimal control method \cite{glaser_training_2015}. Following Ref.~\cite{jandura_time-optimal_2022}, we choose the numerical GRAPE algorithm \cite{, khaneja_optimal_2005} that assumes that $\Omega(t)$ is a piecewise constant pulse of duration $\tau$ described by $N\gg 1$ parameters $\Omega_1,...,\Omega_N$ as $\Omega(t)=\Omega_j$ if $t \in [(j-1)\tau/N, j\tau/N]$. For a given set of parameters the cost function $J$ can be found by solving the coupled differential equations $\ket{\dot{\psi}_q^{(0)}} = -iH^{(0)}\ket{\psi_q^{(0)}}$ and $\ket{\dot{\psi}_q^{(1)}} = -iH^{(0)}\ket{\psi_q^{(1)}}-iH^{(1)}\ket{\psi_q^{(0)}}$  \cite{propson_robust_2021}. The pulse minimizing $J$ is found by optimizing over $\Omega_1,...,\Omega_N$ using a gradient descent optimizer, where GRAPE provides an efficient algorithm to calculate the gradient of $J$.

We find that for {\it any} pulse duration $\tau$ longer than a certain critical $\tau_*\approx 14.32/\omax$ there exists a pulse with $J=0$, i.e. a pulse that implements a CZ gate with fidelity $F=1$ and is robust against amplitude deviations.
We refer to the shortest possible pulse with $\tau = \tau_*$ as the ``amplitude-robust'' (AR) pulse. The AR pulse is of the form $\Omega(t) = \omax \exp[i\varphi(t)]$, i.e. it has always maximal amplitude. The laser phase $\varphi(t)$ of the AR pulse as a function of the dimensionless time $t\omax$ is shown in Fig.~\ref{fig:level_scheme_and_pulses}(b), together with the time-optimal (TO) pulse (without any robustness) found in Ref.~\cite{jandura_time-optimal_2022}. We emphasize that the laser phase of the AR pulse is a smooth function of time, which may be easier to implement experimentally than a pulse with discontinuities in the amplitude or phase. The average time spent in the Rydberg state during the AR pulse is $\tau_R =  4.74/\omax$, roughly 60\% longer than the TO pulse, which achieves $\tau_R = 2.96/\omax$.

\begin{figure}
\includegraphics[width=\linewidth]{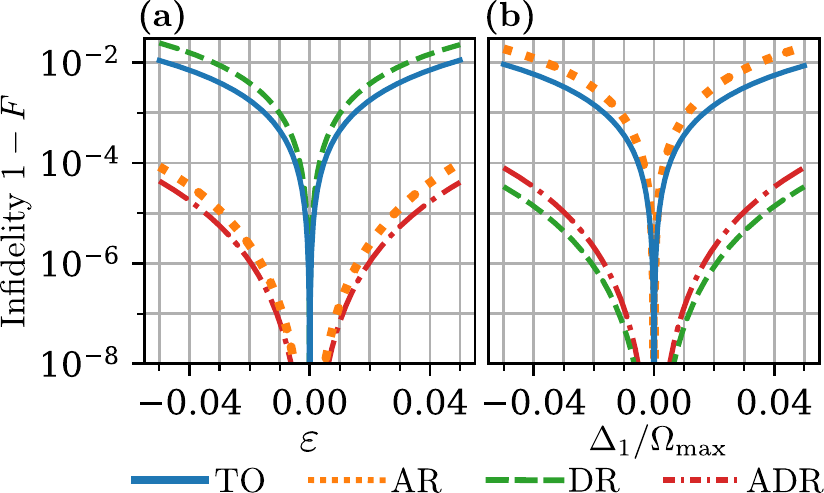}
\caption{\label{fig:single_deviation_plots}
Gate infidelity $1-F$ in the absence of Rydberg decay. (a) Infidelity of the time-optimal (TO), amplitude-robust (AR), Doppler-robust (DR) and amplitude- and Doppler-robust (ADR) pulses as a function of $\varepsilon=\varepsilon_1=\varepsilon_2$, with $\Delta_1 = \Delta_2 = 0$. (b) Infidelity of the same pulses as a function of $\Delta_1$, with $\varepsilon=0$ and $\Delta_2 = 0$.
}
\end{figure}
To demonstrate the robustness of the AR pulse we first calculate the infidelity $1-F$ in the absence of Rydberg decay ($\gamma = 0$), as a function of the amplitude error  $\varepsilon_1=\varepsilon_2=\varepsilon$.
The infidelities are displayed in Fig.~\ref{fig:single_deviation_plots}(a) by the orange dotted (AR pulse) and blue solid  (TO pulse) lines. The AR pulse achieves an infidelity $1-F < 10^{-4}$ even for very large values of $|\varepsilon|$ up to 0.05, improving on the TO pulse by several orders of magnitude. A similar robustness is obtained for $\varepsilon_1 \neq \varepsilon_2$ (not shown).

\subsection{Detuning Robust Pulses}
\label{subsec:detuning_robust_pulses}
Now we turn to pulses which are robust against a detuning of the laser, but not against deviations of the laser amplitude, i.e. we assume $\varepsilon_1=\varepsilon_2=0$. For this setting, we demonstrate analytically that no pulse exists for which the implemented gate is first-order insensitive to $\Delta_1$ and $\Delta_2$. Analogously to the amplitude robust pulse, we expand the Hamiltonians and quantum states as  $H_q = H_q^{(0)} + \Delta_1 H_q^{(1,1)} + \Delta_2 H_q^{(1,2)}$ and $\ket{\psi_q(t)} = \ket{\psi_q^{(0)}} + \Delta_1 \ket{\psi_q^{(1,1)}} + \Delta_2 \ket{\psi_q^{(1,2)}} + \mathcal{O}(\Delta^2)$. Through perturbation theory we find that for any pulse with $\ket{\psi_q^{(0)}} = e^{i\theta_q}\ket{q}$ the first order correction satisfies
\begin{equation}
   \braket{q|\psi_q^{(1,j)}(\tau)} = -ie^{i\theta_q} \int_0^\tau \braket{ \psi_q^{(0)}(t)|H_q^{(1,j)}|\psi_q^{(0)}(t)} \mathrm{d}t.
   \label{eq:first_order_detuning_error}
\end{equation}
By using that $\sum_j H_q^{(1,j)} = (I-\ket{q}\bra{q})$ we see that $\sum_j \braket{q|\psi_q^{(1,j)}(\tau)} = -i\exp(i\theta_q)\tau^{R}_q$,
where \edit{$\tau^{R}_q$}{$\tau^{R}_q = \int_0^\tau \mathrm{d}t (1-|\braket{q|\psi_q^{(0)}}|^2)$} is the \edit{}{average} time spent \edit{}{outside of the computational subspace (i.e.} in the Rydberg state\edit{}{)} when starting in state $\ket{q}$. Since $\tau^{R}_q > 0$ for all pulses which implement a CZ gate, we see that there is no pulse with $\ket{\psi_q^{(1)}(\tau)}=0$. Hence there is no pulse such that the implemented gate is to first order insensitive to $\Delta_1$ and $\Delta_2$. This motivates the search for different solutions to dominant detuning errors in experiments in Sec.~\ref{sec:doppler_robust_pulses} below. Note that the same argument applies even if we restrict the discussion to equal detunings $\Delta_1=\Delta_2=\Delta$. 

We note that while no robust pulse exists, it is still possible to minimize the sensitivity to finite $\Delta$. In Appendix~\ref{app:pulse_most_robust_against_detunings} we use a combination of GRAPE and analytical techniques to find the pulse that minimizes $1-F$ at small but finite values of $\Delta$, while still achieving $F=1$ at $\Delta=0$. This optimal pulse improves the infidelity by only 17\% compared to the TO pulse, an improvement much less relevant than the several orders of magnitude achieved by, e.g., the AR pulse against amplitude deviations or the pulses described below. 

\edit{}{
\subsection{Stark-Shift Robust Pulses}
\label{subsec:stark_shift_robust_pulses}
\begin{figure}
    \centering
    \includegraphics[width=\linewidth]{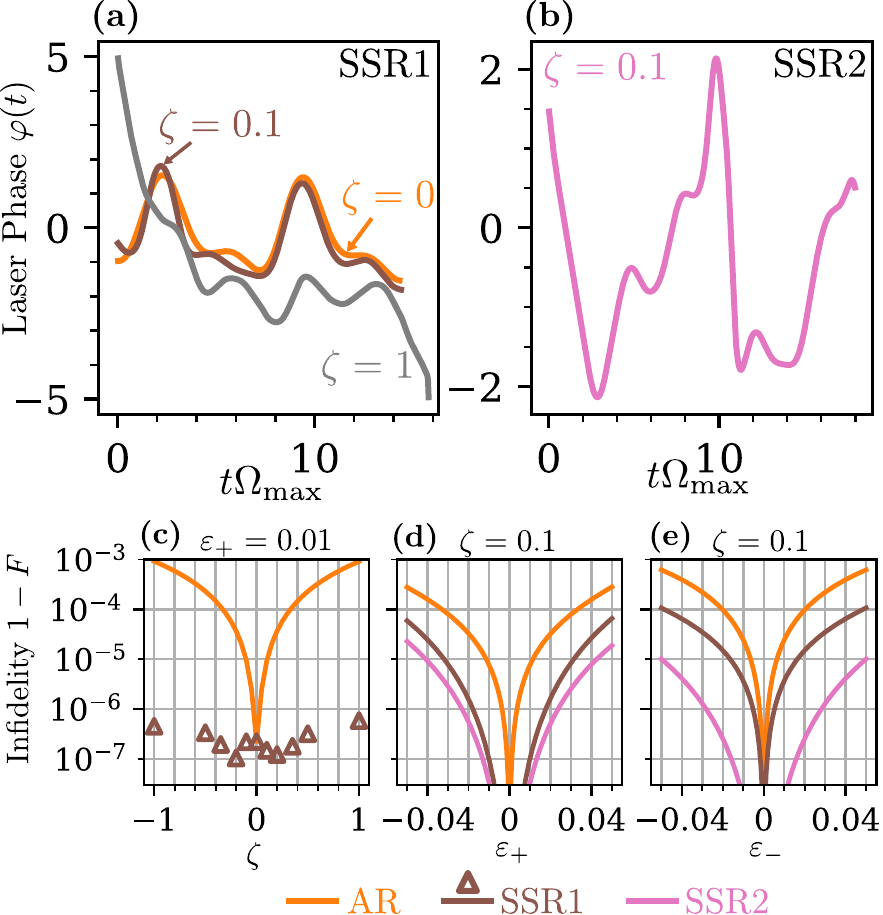}
    \caption{a) The Stark shift robust pulses at identical errors $\varepsilon_1=\varepsilon_2$ (SSR1) for $\zeta = 0.0$ (identical to AR pulse), $\zeta=0.1$ and $\zeta=1$
    b) The Stark shift robust pulses at independent errors $\varepsilon_1\neq=\varepsilon_2$ (SSR2) for $\zeta = 0.1$
    c) The infidelity $1-F$  for the AR pulse (solid lines) and the SSR1 pulses (triangles) as a function of $\zeta$ at $\varepsilon_1=\varepsilon_2=0.01$
    d)(e) The infidelity as a function of $\varepsilon_+$($\varepsilon_-$) for the AR, SSR1 and SSR2 pulse at $\zeta = 0.1$}
    \label{fig:stark_shift_plot}
\end{figure}

One important source of detuning are uncompensated AC-Stark shifts arising due to the off-resonant coupling of the laser to other states. For an uncertain Rabi frequency $(1+\varepsilon_i)\omax$, these Stark shifts are of the form $\chi (1+\varepsilon_i)^2\omax^2$. The known part $\chi \omax^2$ of the Stark shift can be compensated by adjusting the laser frequency accordingly, leaving an unknown detuning $\Delta_i = \zeta \varepsilon_i\omax + \mathcal{O}(\varepsilon^2)$, where $\zeta = 2\chi \omax$ is a dimensionless quantity measuring the strength of the Stark shift. 

Crucially, the detuning $\Delta$ induced by a Stark shift   and the amplitude deviation $\varepsilon$ are \emph{correlated}. This allows, in contrast to Sec.~\ref{subsec:detuning_robust_pulses}, for the existence of pulses which are to first order robust against Stark shifts, so called Stark shift robust (SSR) pulses. For SSR pulses we distinguish between the case of identical errors on both atoms ($\varepsilon_1=\varepsilon_2)$, and the case of independent errors ($\varepsilon_1 \neq \varepsilon_2)$. For the case of identical errors, the SSR pulse (termed SSR1) can be found analogously to the AR pulse (Sec.~\ref{subsec:amp_robust_pulses}) by changing the first order contribution $H_q^{(1)}$ of the Hamiltonian to include the Stark shift (e.g. $H_{01} = \Omega/2\ket{10}\bra{r0} + \hc + \zeta \omax \ket{r0}\bra{r0}$). The SSR1 pulses for $\zeta=0.1$ and $\zeta=1$ are shown in Fig.~\ref{fig:stark_shift_plot}(a) together with the AR pulse ($\zeta=0$). The shape of the SSR1 pulse is a small perturbation of the AR pulse for small $\zeta$ (see $\zeta=0.1$), while for large $\zeta$ its shape is qualitatively different from the AR pulse (see $\zeta=1$). The SSR1 pulse for $\zeta=0.1(\zeta=1)$ spends an average time of $\tau_R = 4.76/\omax$ ($\tau_R = 4.22/\omax$) in the Rydberg state , comparable to the AR pulse. For $\zeta \gtrsim 2$, the optimization procedure fails to find an SSR1 pulse, which is consistent with the fact that for a pure detuning error ($\zeta \rightarrow \infty$), no robust pulse exists (Sec.~\ref{subsec:detuning_robust_pulses}).

To quantify the performance of the SSR1 pulse compared to the AR pulse, Fig.~\ref{fig:stark_shift_plot}(c) shows the infidelity $1-F$ at $\varepsilon_1=\varepsilon_2 = 0.01$ for different values of $\zeta$ for the AR pulse (orange line) and the SSR1 pulses (brown triangles). While the infidelity of the AR pulse strongly increases with increasing $\zeta$, the infidelity of the SSR pulses stays constant and outperforms the AR pulse at $|\zeta|=1$ by more than three orders of magnitude. 

For the case of independent errors, it has to be ensured that the final state $\ket{\psi_{11}}$ is also robust against amplitude deviations if $\varepsilon_-  = (\varepsilon_1-\varepsilon_2)/2\neq 0$. This is achieved by expanding $H_{11} = H_{11}^{(0)} + \varepsilon_+ H_{11}^{(1)} + \varepsilon_- H_{\overline{11}}^{(1)}$, where $H_{\overline{11}}^{(1)} = \zeta \omax \ket{W_+}\bra{W_-}+\hc$ contains only Stark shift terms, and including the corresponding $\braket{\psi_{\overline{11}}^{(1)}|\psi_{\overline{11}}^{(1)}}$ term in the cost function~\eqref{eq:cost_function}. The resulting SSR pulse (termed SSR2) is displayed in Fig.~\ref{fig:stark_shift_plot}(b) for $\zeta=0.1$. In contrast to the SSR1 pulse it is qualitatively different from the ADR pulse, due to the additional requirement that $\ket{\psi_{\overline{11}}^{(1)}}=0$. The SSR2 pulse for $\zeta=0.1$ spends an average time of $\tau_R = 5.87/\omax$ in the Rydberg state, roughly 25\% more than the SSR1 pulse.

The performance difference between the SSR1 and SSR2 pulse is demonstrated in Fig.~\ref{fig:stark_shift_plot}(d,e). Panel (d) shows the infidelity of the AR (orange), SSR1(brown) and SSR2(pink) pulse at $\zeta=0.1$ as a function of $\varepsilon_+$, while $\varepsilon_-=(\varepsilon_1-\varepsilon_2)/2=0$. Here, the SSR1 and SSR2 pulses show a similar infidelity, and both significantly outperform the AR pulse. In contrast, panel (e) shows the infidelity of the same pulses as a function of $\varepsilon_-$, while $\varepsilon_+=0$. As expected, the SSR2 pulse now significantly outperforms both the AR and the SSR1 pulse.  
}

\section{Doppler Robust Pulses}
\label{sec:doppler_robust_pulses}
\begin{figure}
\includegraphics[width=\linewidth]{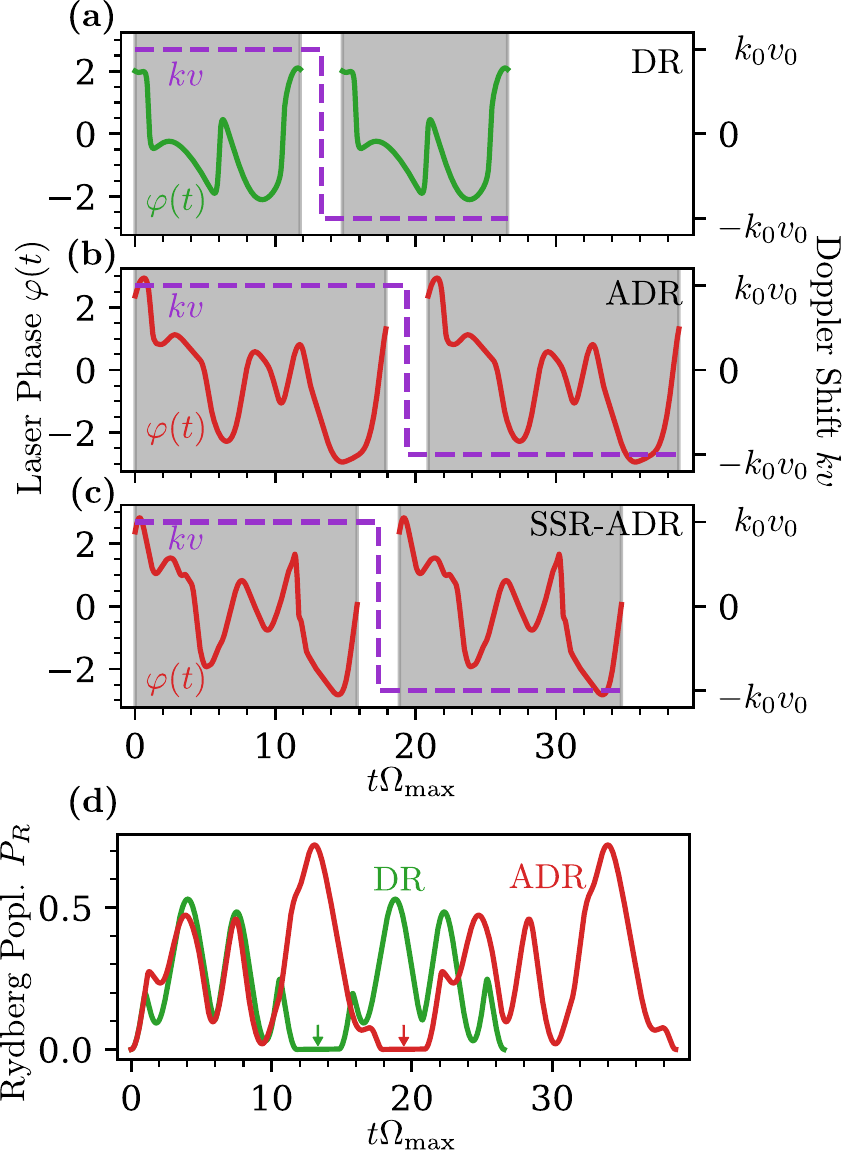}
\caption{\label{fig:doppler_robust_pulses_and_potential} Laser phase for a) the  Doppler-robust (DR, green) b) the amplitude- and Doppler-robust (ADR, red), and c) the Stark-shift robust (SSR) ADR pulse . Each pulse consists of two identical halves, shown by the grey areas, applied with opposite Doppler shifts $\Delta = kv$ (purple dashed lines), achieved as described in the text. In each of the halves the laser amplitude is maximal ($|\Omega(t)|=\omax$), while $\Omega(t)=0$ outside of the grey areas.
d) The population $P_R$ of the Rydberg state (averaged over the four computational basis states as initial states) for the DR and ADR pulse as a function of time. Note that between the two halves of the pulses (shown by the arrows) the population of the Rydberg state is zero.}
\end{figure}

\edit{The practically most relevant}{A second practically relevant} source of detuning error is the Doppler shift $\Delta_j = kv_j$ where $k$ is the wavevector of the laser and $v_j$ is the velocity of atom $j$ along the direction of the laser. In contrast to a fixed detuning discussed in Sec.~\ref{subsec:detuning_robust_pulses}, the sign of the Doppler shift can be flipped by changing the sign of $k$ or $v$. In the following we will argue that a robust gate can be achieved by splitting the gate into two identical halves applied sequentially, with the sign of $\Delta_j$ reversed between the two halves, as illustrated in Fig.~\ref{fig:doppler_robust_pulses_and_potential}(a). We start in Sec.~\ref{subsec:design_of_doppler_robust_pulses} by showing how this reversal of $\Delta_j$ allows for pulses robust against Doppler shifts. In  Sec.~\ref{subsec:switching_the_doppler_detuning} we then discuss two potential experimental methods for reversing $\Delta_j$.

\subsection{Design of Doppler Robust Pulses}
\label{subsec:design_of_doppler_robust_pulses}

In order to be robust against Doppler errors, we use GRAPE to search for a pulse $\Omega(t)$ of duration $\tau$ that satisfies two conditions. First, implementing a controlled-$R_z(\pi/2)$ gate when $\Delta=0$ (i.e., satisfying $\ket{\psi_q^{(0)}(\tau)} = e^{i\theta_q}\ket{q}$ with $\theta_{11}-\theta_{10}-\theta_{01} = (2n+1/2)\pi$ for an integer $n$). Second, achieving a first order error $\ket{\psi_q^{(1,j)}}$ which is entirely along the direction of $\ket{q}$, i.e. $(I-\ket{q}\bra{q})\ket{\psi_q^{(1,j)}(\tau)} = 0$. The state after applying this pulse once is then
\begin{equation}
\ket{\psi_q(\tau)} = \left(e^{i\theta_q}+\sum_j\Delta_j \braket{q|\psi_q^{(1,j)}(\tau)}\right)\ket{q} + \mathcal{O}\left(\Delta^2\right).
\end{equation}
When applying  the pulse $\Omega(t)$ a second time, the sign of $\Delta_j$ is reversed. This implies $\ket{\psi_q(2\tau)} = e^{2i\theta_q}\ket{q}+\mathcal{O}(\Delta^2)$, and thus the combined pulse is robust against Doppler errors. Crucially, the Rydberg state population after the first pulse is of order $\mathcal{O}(\Delta^2)$, which makes the gate insensitive to the relative phase of the lasers between the two pulses and also allows an arbitrary waiting time between the pulses without incurring errors from Rydberg state decay. 

GRAPE can be applied to this problem analogously to the AR-case, with the cost function
\begin{equation}
    J = 1-F + \sum_{j,q} \braket{\psi_q^{(1,j)}(\tau)|(I-\ket{q}\bra{q})|\psi_q^{(1,j)}(\tau)}.
    \label{eq:cost_function_doppler}
\end{equation}
The shortest possible pulse which is robust against Doppler errors, called the ``Doppler-robust'' (DR) pulse, is shown in Fig.~\ref{fig:doppler_robust_pulses_and_potential}(a). The population of the Rydberg state (averaged over the four computational basis states as initial states) in shown in Fig.~\ref{fig:doppler_robust_pulses_and_potential}(d). As mentioned above, the population of the Rydberg state vanishes between the two pulses. The average time that the DR pulse spends in the Rydberg state (over the entire gate) is given by $\tau_R = 5.56/\omax$. 

By simply adding the cost functions for the AR and the DR cases, we can  identify the shortest possible pulse which is robust against both imperfections, which we call the ``amplitude- and Doppler-robust'' (ADR) pulse. The laser phase of the ADR pulse is displayed in Fig.~\ref{fig:doppler_robust_pulses_and_potential}(b), the population of the Rydberg state in Fig.~~\ref{fig:doppler_robust_pulses_and_potential}(d). The ADR pulse spends an average time \edit{$\tau_R = 13.20/\omax$}{$\tau_R = 10.37$} in the Rydberg state, and is thus significantly more affected by its decay than the other three pulses.  We remark that the ADR pulse is also robust against amplitude deviations $\varepsilon_i$ that are different in the two halves of the pulse, since each half is individually robust against amplitude deviations.

The infidelity of all four pulses (TO, AR, DR and ADR) as a function of the detuning $\Delta_1$ of the first atom is shown in Fig.~\ref{fig:single_deviation_plots}(b). For the TO and AR pulse the detuning is kept constant, while for the DR and ADR pulse its sign is switched after the first half of the pulse. The DR and ADR pulse achieve $1-F < 10^{-4}$ for $|\Delta_1|/\omax < 0.05$, two to three orders of magnitude better than the TO and AR pulses. We also compare the performance of the DR and ADR pulses to the TO and AR pulses when varying the amplitude deviation $\epsilon$ in Fig.~\ref{fig:single_deviation_plots}(a). As expected, the DR pulse does not show any robustness to amplitude deviations and behaves similar to the TO pulse, while the ADR pulse outperforms not only the TO pulse, but also the AR pulse.

\edit{}{Analogously to the AR pulse, a Stark shift robust version exists also for the ADR pulse, shown for $\zeta=0.1$ in Fig.~\ref{fig:doppler_robust_pulses_and_potential}(d). As for the AR pulse, the Stark shift robust ADR pulse is qualitatively similar to the ADR pulse in the absence of a Stark shift. It spends an average time of $\tau_R = 10.66/\omax$ in the Rydberg state. Note that  the ADR pulses is inherently robust against Stark shift errors with $\varepsilon_1=-\varepsilon_2$, so  that in contrast to the AR pulse no distinction between identical and independent errors is necessary. To see this, first note that for the states $\ket{\psi_{10}}$ and $\ket{\psi_{10}}$ the distinction between identical and independent errors is irrelevant, because only one of the two atoms is affected by the error. The remaining state $\ket{\psi_{\overline{11}}}$ (describing the state of the atoms \emph{after the first pulse half} when staring in $\ket{11}$) is intrinsically robust against Stark shift errors with $\varepsilon_1 = - \varepsilon_2$, because these errors only result in a nonzero detuning $\Delta_- = \zeta\varepsilon_1|\omax|$ and a corresponding perturbation $H_{\overline{11}}^{(1)} = \zeta \omax \ket{W_+}\bra{W_-}+\hc$. By the Doppler robustness of the ADR pulse, it holds that $\braket{W_\pm|\psi_{\overline{11}}^{(1)}}=0$. But because $H_{\overline{11}}^{(1)}$ only leads to the population of $\ket{W_-}$ and otherwise leaves the evolution unchanged, it also holds that $\braket{11|\psi_{\overline{11}}^{(1)}}=0$, so that $\ket{\psi_{\overline{11}}^{(1)}}=0$. Hence each of the two pulse halves of the ADR pulse are robust against Stark shift errors with $\varepsilon_1 = - \varepsilon_2$, so also the whole ADR pulse is robust against those errors. Note that this even holds if the $\varepsilon_i$ are different in each of the pulse halves. }

\subsection{Reversing the Doppler Shift}
\label{subsec:switching_the_doppler_detuning}
The DR and ADR pulse require that $\Delta_j$ is reversed after the first half of the pulse. Here we propose two methods for switching the sign of $\Delta_j$, called the \emph{switch} method and the \emph{wait} method.

In the switch method, the direction of the laser (i.e. the sign of $k$) is reversed between the two pulses. Note that the switch method works regardless of the relative phase between the two laser beams, because the Rydberg population in between the two pulse halves vanishes [see Fig.~\ref{fig:doppler_robust_pulses_and_potential}(b)].

The wait method instead makes use of the fact that the atom is confined in a potential that is approximately harmonic, and therefore the velocity in the direction of the laser propagation is periodic with trap frequency $\otr$. By  waiting for a time $\pi/\otr$ between the two pulses, the sign of $v$, and thus the sign of $\Delta_j$, is reversed. Since the Rydberg state is not populated between the two pulse halves, no additional decay error arises during this wait time even if $\otr \ll \gamma$.

The wait method makes several implicit assumptions on the motion of the atoms. First, we assume that the propagation direction of the laser is along one of the normal modes of the trap. Second, we assume that the coherence of the atomic motion is much longer than one motional period \cite{kuhr_analysis_2005}. Finally, we assume that the atomic temperature is low enough for the trap anhamonicity to be negligible. In Appendix~\ref{subsec:robustness_to_trap_inhomogeneities_and_anharmonicities} we estimate the impact of a finite trap anharmonicity and show that for achievable experimental parameters it does not significantly affect the gate performance.

Both the switch and the wait method require that the velocity of the atoms is approximately constant during each pulse. The acceleration of the atoms during the pulses is thus a source of error. For the switch method this error can be avoided by abruptly turning off the trapping potential during the pulses, which is already common practice in many experiments to avoid differential light shifts and anti-trapping of the Rydberg state \cite{saffman_quantum_2016}. However, this approach is unsuitable for the wait method, since it affects the velocity reversal, and is also undesirable because it heats the atom and may prevent the execution of deep circuits with many gates. To mitigate these disadvantages, we propose to modulate the trapping potential sinusoidally in time, and to apply the pulses at times where the potential is zero (Appendix~\ref{app:modulation_of_the_trap_frequency}). This gives rise to an approximately constant velocity of the atoms during the pulses while also eliminating the differential light shift and heating from square-wave modulation \cite{tiecke_nanophotonic_2014}. For the wait method, we show  that it is possible to apply the two pulse halves at two different times of vanishing potential such that the velocity is reversed between the two pulses. Note that fast sinusoidal trap modulation was experimentally demonstrated in an optical tweezer for the purpose of eliminating light shifts in a cavity QED experiment \cite{tiecke_nanophotonic_2014}. For the remainder of this work we assume that the trap modulation is applied for both the switch and the wait method, in Appendix~\ref{subsec:performance_without_trap_modulation} we discuss the two methods \emph{without} the trap modulation.

In the last two subsections we have shown how the reversal of the Doppler shift in the middle of the gate allows to find a pulse that is robust against errors arising from Doppler shifts, and a pulse which is robust against both amplitude deviations and Doppler shifts. We demonstrated that the infidelity arising due to Doppler shifts is reduced by several orders of magnitude by the robust pulses, and provided two methods to switch the sign of the Doppler shift.

\section{Infidelities in a Realistic Error Model}
\label{sec:infidelities_in_a_realistic_error_model}

\begin{figure}
\includegraphics[width=\linewidth]{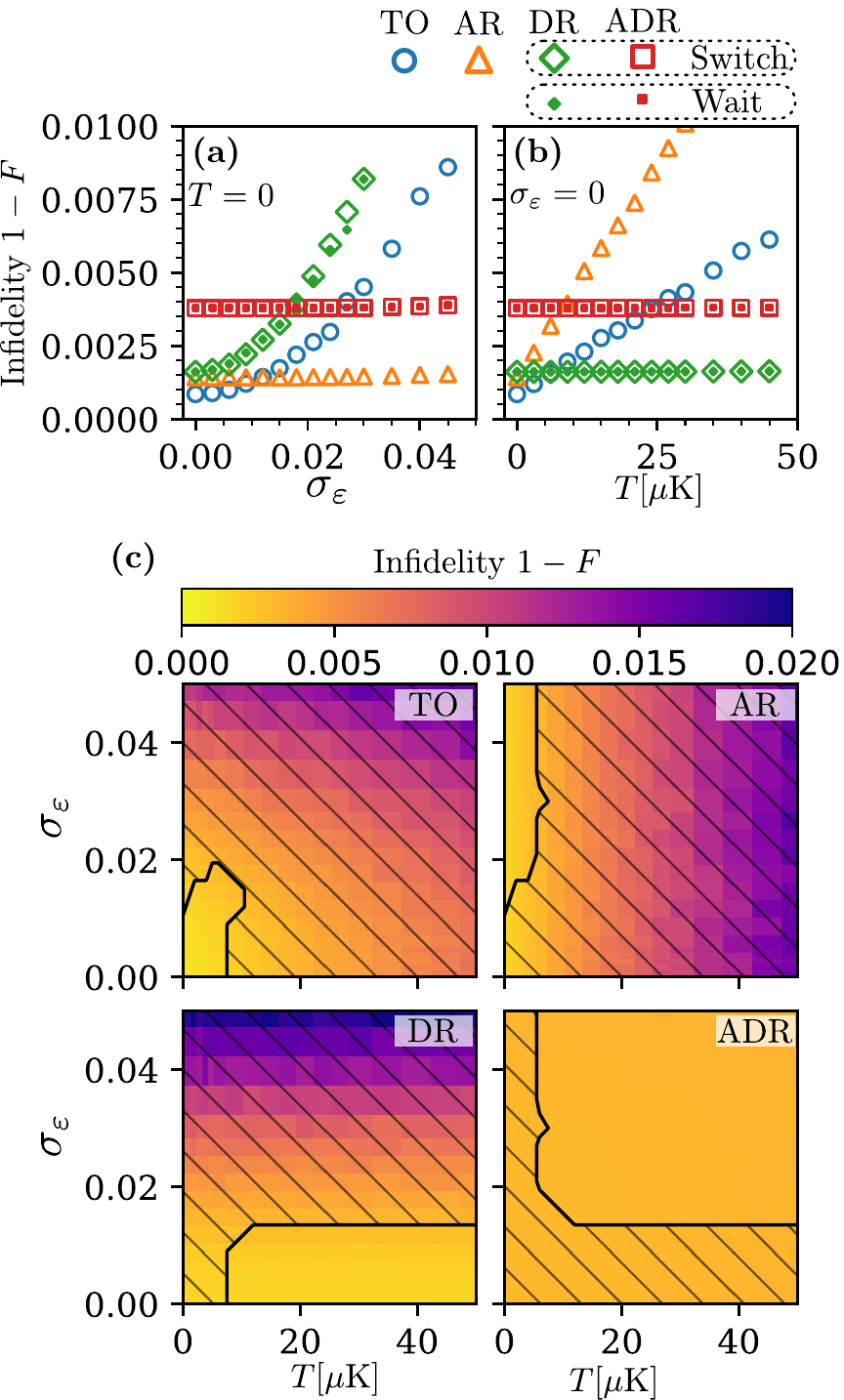}
\caption{The infidelity $1-F$ of the TO, AR, DR and ADR pulse at different values of the amplitude uncertainty $\sigma_\varepsilon$ and the atomic temperature $T$. a) $1-F$ as a function of $\sigma_\varepsilon$ at $T=0$. For the DR and ADR pulse, open symbols show the infidelity with the switch method, while filled symbols show the infidelity with the wait method. Sinusoidal modulation of the trap is applied in all cases. b) $1-F$ as a function of $T$ at $\sigma_\varepsilon=0$. c) Color plot of $1-F$ as a function of both $\sigma_\varepsilon$ and $T$. For the DR and ADR pulse we use the wait method including the sinusoidal modulation of the trap. The encircled regions show the range of imperfections where each pulse performs the best.
}
\label{fig:unconditional_fidelity}
\end{figure}

To assess the performance of these pulses in a realistic experiment, we now include the decay of the Rydberg state and atomic motion in a harmonic trap. For this we assume random and uncorrelated initial velocities and positions for both atoms, drawn from a normal distribution with standard deviation $\sqrt{k_BT/m}$ and $\sqrt{k_BT/m\otr^2}$, respectively, where $T$ is the temperature and $m$ is the mass of the atoms. We then assume that the atoms follow classical trajectories in the harmonic trap, which we incorporate as a modified Rabi frequency $\tilde{\Omega}(t) = e^{-ikx(t)}\Omega(t)$. In the case of the DR and ADR pulses, we simulate both the switch and wait method of reversing the detuning between the two pulse halves, applying the modulation of the trapping potential as described in Appendix~\ref{app:modulation_of_the_trap_frequency} for both methods.
We sample the laser amplitude error $\varepsilon$ from a normal distribution with standard deviation $\sigma_\varepsilon$. For simplicity, we set $\varepsilon_1=\varepsilon_2=\varepsilon$, but note that the same robustness is achieved if $\varepsilon_1 \neq \varepsilon_2$ since the AR and ADR pulses are robust against $\varepsilon_1$ and $\varepsilon_2$ independently. 

For the Rydberg excitation, we consider parameters recently proposed for metastable $~^{171}$Yb qubits using a single-photon excitation to the $\ket{75~^3S_1\,F=3/2}$ Rydberg state \cite{wu_erasure_2022}, although we note that these are broadly similar to proposed or achieved values for other alkaline earth atoms such as Sr \cite{madjarov_high-fidelity_2020,pagano_error_2022} and ground-state $~^{171}$Yb qubits \cite{ma_universal_2022,jenkins_ytterbium_2022}. The specific numerical values considered here are: $\omax = 2\pi \times 5.5$~MHz, $2\pi/k =302$~nm,  $1/\gamma = 100\,\mu$s, $\otr = 2\pi \times 50$~kHz and $m = 171$~u. The blockade strength is $B \sim 5 \text{ THz}\,\mu\text{m}^6/r^6$, so that $B\gg \omax$ for realistic values of $r$ in the range of 3-6 $\mu$m \cite{ma_universal_2022}. In the following calculations we assume a perfect Rydberg blockade.Similar parameters can be obtained for alkali atoms, but we note that two-photon excitation typically reduces the wavevector associated with Doppler shifts, at the expense of an additional decay error from the decay of the intermediate state. \edit{}{As the Stark shift strength in metastable $^{171}$Yb qubits is unknown, we assume the value measured in $^{88}$Sr qubits, given by $2\pi\chi \approx 10 \text{kHz}/\text{MHz}^2$ \cite{madjarov_high-fidelity_2020}.  For the Rabi frequency $\omax$ this corresponds to $\zeta=0.1$ . In the following we always use the Stark shift robust variants of the AR and ADR pulse. Since we restrict ourselves to the $\varepsilon_1=\varepsilon_2$ case, we use the SSR1 pulse as the Stark shift robust variant of the AR pulse.} 

The results are summarized in Fig.~\ref{fig:unconditional_fidelity}. We first consider the performance with only amplitude or Doppler errors. The infidelity as a function of $\sigma_\varepsilon$ with $T=0$  is shown in Fig.~\ref{fig:unconditional_fidelity}(a). For small values of $\sigma_\varepsilon$ the decay of the Rydberg state is the dominant error, so the fidelities depend only on the time spent in the Rydberg state. The order of the pulses by increasing time spent in the Rydberg state (in our case identical to increasing pulse duration) is TO, AR, DR, ADR. In contrast, as $\sigma_\varepsilon$ increases, the infidelity of the AR and ADR pulses stays almost constant, while the infidelity of the TO and DR pulses increases quadratically. At  $\sigma_\varepsilon \gtrsim 0.010$ the AR pulse becomes favourable compared to the TO pulse, at $\sigma_\varepsilon \gtrsim 0.026$ the ADR pulse becomes favorable compared to the TO pulse. The AR pulse outperforms the ADR pulse, because while both pulses are robust to deviations of the laser amplitude, the AR pulse spends less time in the Rydberg state.

Next we consider the performance as a function of $T$, in the absence of amplitude errors ($\sigma_\varepsilon=0$) [Fig.~\ref{fig:unconditional_fidelity}(b)]. At low temperatures the pulses are again ordered by the time they spent in the Rydberg state. With increasing temperature, however, the infidelity of the DR and ADR pulses stays almost constant, while the infidelity of the TO and AR pulses increases linearly with $T$ (quadratically with $\Delta$). For $T\gtrsim 6\mu$K the DR pulse outperforms the TO pulse, for $T\gtrsim 28\mu$K the ADR pulse outperforms the TO pulse. These results do not depend on whether the switch or the wait method is used. We note that the infidelity of the TO pulse at elevated temperatures is roughly consistent with previous estimates for various non-robust blockade gates \cite{saffman_quantum_2016,pagano_error_2022}. We remark that with increasing temperature, other imperfections not considered here, such as the anharmonicity of the trap may become increasingly relevant (see Appendix~\ref{subsec:robustness_to_trap_inhomogeneities_and_anharmonicities} for a discussion).

Finally, we consider the infidelity in the presence of both amplitude and Doppler errors. Fig.~\ref{fig:unconditional_fidelity}(c) shows the infidelity of all four pulses over a range of imperfections. Additionally, the region in which each pulse has the lowest infidelity out of the four considered pulses is marked. The results are shown for the wait method, we verified that for the switch method identical results are obtained. As expected the TO pulse performs best when all imperfections are small, while ADR pulse is the best pulse for large amplitude uncertainties and large temperatures. The AR and DR pulses are the best choice when either the amplitude uncertainty is large or the temperature is large, while the other imperfection is small.

\section{Conditional Infidelity and Logical Error Rate}
\label{sec:conditional_infidelity_and:logical_error_rate}

\begin{figure*}
\includegraphics[width=\linewidth]{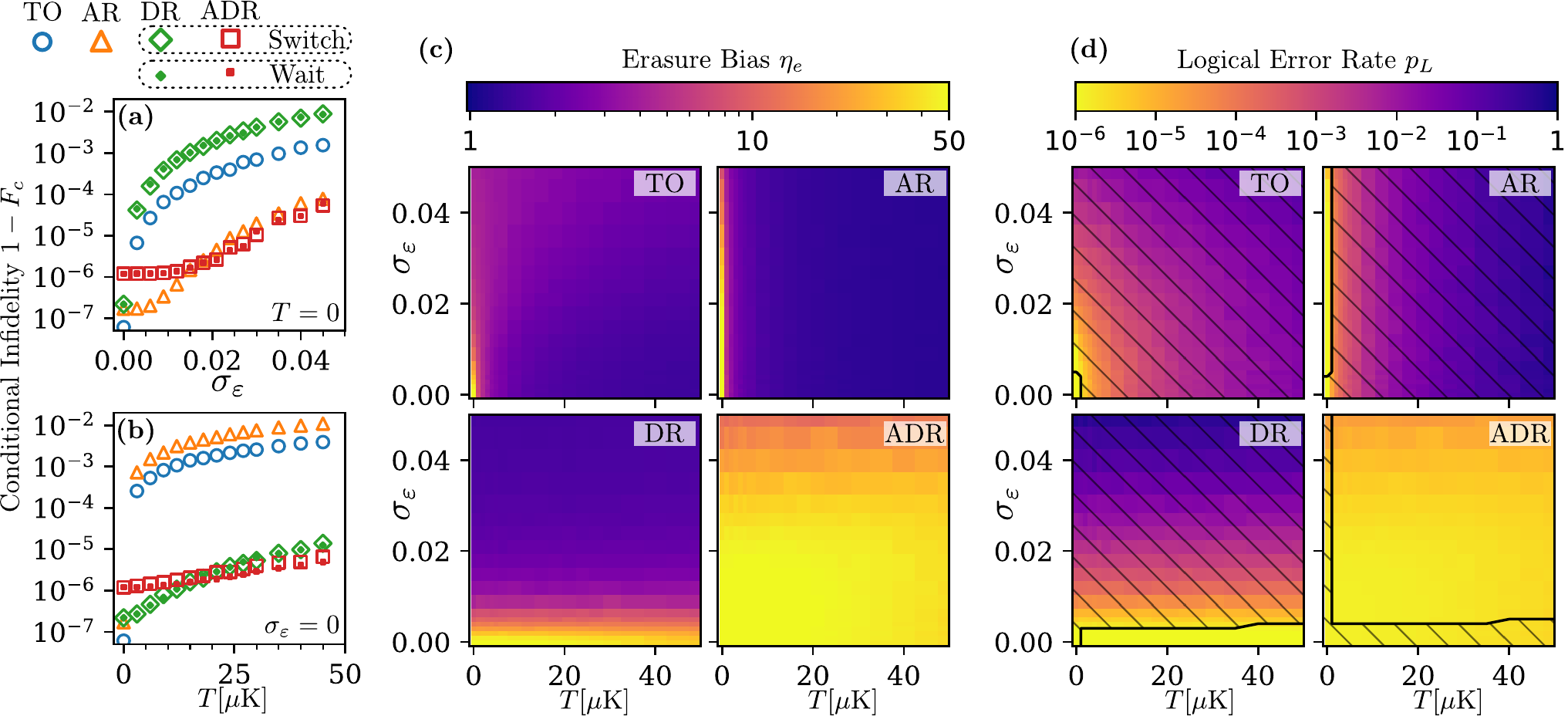}
\caption{a) Conditional fidelity $1-F_c$ of the four pulses as a function of the amplitude uncertainty $\sigma_\varepsilon$ at $T=0$. For the DR and ADR pulse, open symbols show results with the switch method, while filled symbols show the results with the wait method. For both methods the sinusoidal modulation of the trap is applied. b) $1-F_c$ as function of $T$ at $\sigma_\varepsilon=0$.  c) The erasure bias $\eta_e$ as a function of $T$ and $\sigma_\varepsilon$ (using the wait method including the trap modulation for the DR and ADR pulse), assuming $\eta_e=50$ in the absence of imperfections. d) The logical error rate $p_L$ as a function of $T$ and $\sigma_\varepsilon$. The encircled regions show the range of imperfections where each pulse performs the best.
}
\label{fig:conditional_fidelity}
\end{figure*}

In the context of FTQC, both the error probability and the type of error are important. Recently, it was proposed that using the metastable state in $~^{171}$Yb to encode qubits ensures that the vast majority of Rydberg decay errors result in transitions out of the computational subspace that can be efficiently detected \cite{wu_erasure_2022}. This converts decay errors into erasure errors, for which the FTQC threshold is much higher.

Maintaining this advantage in the presence of experimental imperfections requires that the fraction of errors converted into erasures, $R_e$, is close to unity. From the fact that a small fraction of decays of the Rydberg state does lead back to the computational subspace, Ref.~\cite{wu_erasure_2022} estimated that $R_e=0.98$ for the case of spontaneous decays from the Rydberg state, which is the only fundamental limitation to the fidelity of multi-qubit Rydberg gates. In this case, the estimated XZZX surface code threshold can be as high as $p_{th} = 4.15\%$, compared to $p_{th} = 0.93\%$ for a comparable Pauli error model \cite{wu_erasure_2022}. Similar estimates have been made for other erasure-biased qubits \cite{kubica_erasure_2022}.

To understand the error decomposition of our robust pulses, we start by assuming that \emph{all} Rydberg decay errors lead to transitions outside of the computational subspace and compute the \emph{conditional fidelity} $F_c$, i.e. the fidelity conditioned on the final state being in the computational subspace. The conditional infidelity $1-F_c$ is shown as a function of the amplitude uncertainty $\sigma_\varepsilon$  at $T=0$ in Fig.~\ref{fig:conditional_fidelity}(a) and as a function of $T$ at $\sigma_\varepsilon=0$ in Fig.~\ref{fig:conditional_fidelity}(b), assuming the same experimental parameters as in Sec.~\ref{sec:infidelities_in_a_realistic_error_model}. The robust pulses always outperform the non-robust pulses by several orders of magnitude and satisfy  $1-F_c \ll 1-F$ [see Fig.~\ref{fig:unconditional_fidelity}(a),(b) and Fig.~\ref{fig:conditional_fidelity}(a),(b)], showing that errors are dominated by transitions out of the computational subspace. This is expected because the robust pulses effectively trade sensitivity to imperfections for Rydberg decay. Note that again the switch and the wait method give identical results in Figs.~\ref{fig:conditional_fidelity}(a)(b). For the rest of this work we show the results from the wait method, but have verified that they do not change significantly when using the switch method instead.

To quantify the logical error rate achievable for the robust pulses for metastable $~^{171}$Yb qubits, we now include that not all, but only a fraction of \edit{$R_e=0.98$}{$r=0.98$} of the decay errors are converted into erasures \cite{wu_erasure_2022}. \edit{}{To calculate the logical error rate we assume, analogously to Ref.~\cite{wu_erasure_2022}, a Pauli error channel, in which an erasure error occurs with probability $p_e$ and a random Pauli error occurs with probability $p_p$. 
We identify $p_e$ and $p_p$ from the probability $p_d$ of a decay error and the conditional infidelity $1-F_c$ in the exact error model as $p_e = rp_d$ and $p_p \approx (1-r)p_d+(1-p_d)(1-F_c)$. Then a fraction $R_e = p_e/(p_e+p_d)$ of all errors are converted into erasures.} 
For the pulse proposed here, we compute a quantity related to $R_e$, the erasure bias \edit{$\eta_e = 1/(1-R_e) = (1-F)/(1-F_c)$}{$\eta_e = 1/(1-R_e)$}. A larger erasure bias implies a larger fraction of erasure errors, and thus a higher threshold error probability. \edit{Note that the value of $1-F_c$ here is larger than in Fig.~\ref{fig:conditional_fidelity}(a),(b), since it also contains a contribution from decay errors that lead back to the computational subspace.}{} In the $\varepsilon=\Delta=0$ case the erasure bias is then given by $\eta_e=50$, the predicted maximum value for $~^{171}$Yb \cite{wu_erasure_2022}.  Inspecting Fig.~\ref{fig:conditional_fidelity}(c), it is clear that the TO pulse only achieves a large value of $\eta_e$ for very small temperatures and amplitude uncertainties and that $\eta_e$ drops rapidly to $\sim 1$ in the presence of significant amplitude or Doppler errors. The ADR pulse instead maintains $\eta_e \approx 50$ over the whole range of considered parameters of $\sigma_\varepsilon$ up to 0.05 and $T$ up to 50$\mu$K. The DR pulse achieves $\eta_e\approx 50$ as long as $\sigma_\varepsilon$ is small, but $\eta_e$ drops rapidly as $\sigma_\varepsilon$ increases, while the AR pulse shows the opposite behavior. 

Using the total infidelity and $\eta_e$, we can estimate the logical error rate for a given error correcting code. For concreteness, we consider the $d=5$ XZZX surface code, where the logical error rate after a single round of fault-tolerant error correction, $p_L$, is presented in Ref.~\cite{wu_erasure_2022} for a range of physical error rates $p=1-F$ and  erasure biases $\eta_e$. The estimated value of $p_L$ is shown in Fig.~\ref{fig:conditional_fidelity}(d) for the four pulses studied here. Remarkably, the ADR gate outperforms the other three gates by many orders of magnitude unless either $\sigma_e$ or $T$ is very small. This is in contrast to Fig.~\ref{fig:unconditional_fidelity}(c), which shows that the total infidelity of the ADR gate is higher than the other three sequences until $\sigma_e$ or $T$ reach modest values. This illustrates a fundamental tradeoff that is one of the central results of this paper: the larger infidelity associated with the longer ADR sequence is more than compensated by the increased $\eta_e$.

\section{Conditionally Robust Pulses}
\label{sec:conditionally_robust_pulses}
\begin{figure}
    \centering
    \includegraphics[width = \linewidth]{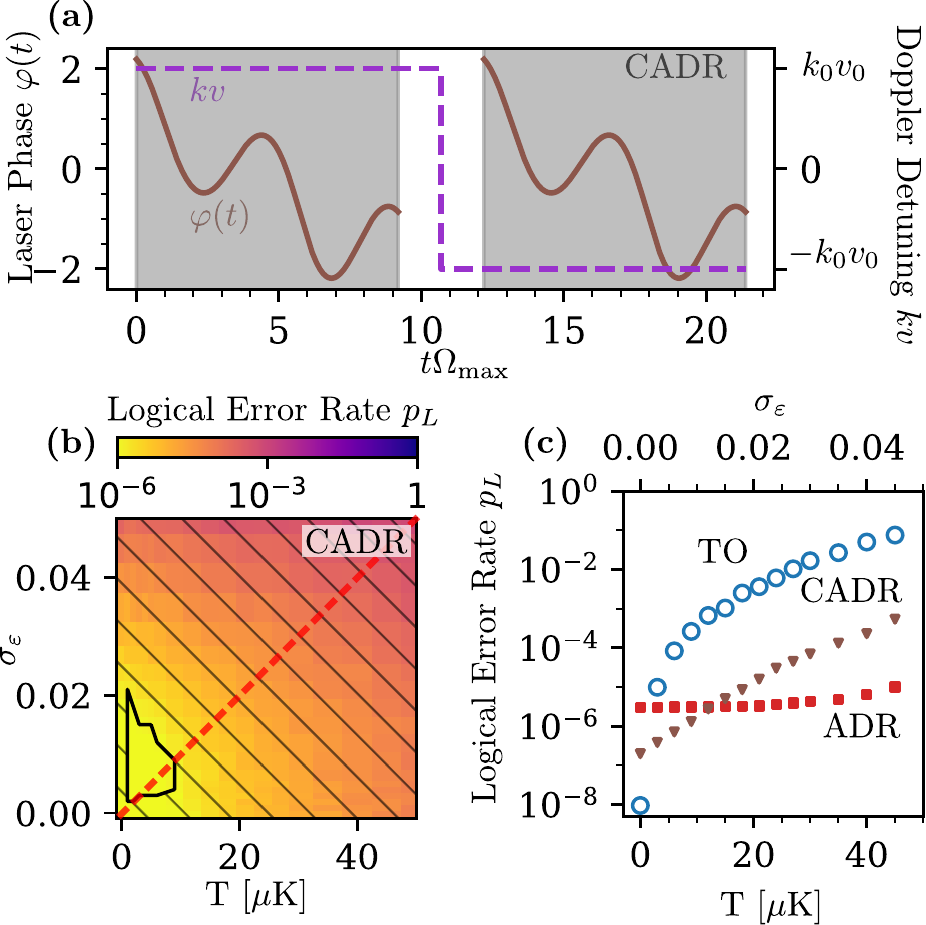}
    \caption{a) The laser phase as a function of time for the conditionally amplitude and Doppler robust (CADR) pulse (brown, solid line). The laser amplitude is maximal whenever the laser is turned on (marked by the gray areas) and zero in between the two halves of the pulse. The Doppler shift $kv$ (purple, dashed line) has to switch sign between the two pulses. b) The logical error rate of the CADR pulse. The encircled region shows the range of imperfections where the CADR pulse outperforms the TO, AR, DR and ADR pulses. c) The logical error rate of the TO, ADR and CADR pulse along the dashed red line in panel b). Below 10 $\mu$K the CADR pulse has a lower logical error rate than the ADR pulse, the improvement is up to one order of magnitude.}
    \label{fig:cadr_plot}
\end{figure}
In Sec.~\ref{sec:conditional_infidelity_and:logical_error_rate} we saw that not only the infidelity, but also the conditional infidelity, plays an important role for the logical error rate. It is thus interesting to design pulses that are not robust, in the sense that the final state is to first order insensitive to imperfections, but only \emph{conditionally} robust, in the sense that only the projection of the final state onto the computational subspace is to first order insensitive to imperfections. The advantage of these conditionally robust pulses is that they can be shorter and less susceptible to decay errors than their unconditional counterparts, since they have fewer robustness constraints to satisfy.

Such conditionally robust pulses can be found with GRAPE by simply replacing the first order error $\ket{\psi_q^{(1)}}$ by its projection $\braket{q|\psi_q^{(1)}}\ket{q}$ onto the computational subspace in the objective function $J$ [see Eqs.~\eqref{eq:cost_function},\eqref{eq:cost_function_doppler}], giving
\begin{equation}
    J = 1-F + \sum_q |\braket{q|\psi_q^{(1)}}|^2
\end{equation}
for conditionally amplitude robust pulses and simply $J=1-F$ for conditionally Doppler robust pulses.

In Fig.~\ref{fig:cadr_plot}(a) we show the fastest possible conditionally amplitude- and Doppler-robust (CADR) pulse. This pulse spends only a time of $\tau_R = 6.61/\omax$ in the Rydberg state, less than half of the time of the ADR pulse and only 2.2 times longer than the TO pulse. Surprisingly, the laser phase $\varphi(t)$ has a simpler shape than for the ADR pulse [See Fig.~\ref{fig:doppler_robust_pulses_and_potential}(a)], simplifying the experimental implementation. The logical error rate of the CADR pulse as a function if $\sigma_\varepsilon$ and $T$ is shown in Fig.~\ref{fig:cadr_plot}(b). We observe that there is a window of moderate temperatures up to 10 $\mu$K and moderate amplitude uncertainties up to 2\% where the CADR pulse outperforms the TO, AR, DR and ADR pulse. For very low imperfections the TO, AR or DR pulse are better than the CADR pulse because they spend even less time in the Rydberg state, for large imperfections the ADR pulse is better because the logical error rate depends on the conditional as well as the unconditional fidelity, and for large enough imperfections the unconditional infidelity for the CADR pulse is larger than for the ADR pulse.  The logical error rate along a diagonal cut [red, dashed line in Fig.~\ref{fig:cadr_plot}(b)] is shown in Fig.~\ref{fig:cadr_plot}(c). We observe that the CADR pulse can improve the logical error rate by up to an order of magnitude over the ADR pulse.

We note that the same approach can be used to produce conditionally amplitude robust or conditionally Doppler robust pulses. However, these pulses only improve the logical error rate over a vanishingly small region in parameter space.

\section{Conclusion}
\label{sec:conclusion}
We have presented \edit{four}{several} new laser pulses that implement a CZ gate using a global laser and are robust against amplitude deviations of the laser and Doppler shifts, where the latter is achieved by reversing the sign of the Doppler shift between two halves of the pulse. Our robust pulses strongly suppress errors from amplitude deviations and Doppler shifts, at the cost of a slightly larger error due to decay of the Rydberg state.
Additionally we estimated the logical qubit performance in the context of the erasure-biased metastable $^{171}$Yb qubit, and found that two of the new pulses (ADR and CADR)  outperform all other pulses unless the imperfections are very small, because they maintain the erasure bias even in the presence of imperfections. Robust pulses enable significant gains from quantum error correction even for significantly elevated temperatures and amplitude deviations.

This work significantly relaxes the technical requirements for FTQC with neutral atoms \cite{saffman_quantum_2016} by extending the erasure conversion concept to amplitude and Doppler shift errors \cite{wu_erasure_2022}. The most important of these is the constraint to have near-ground-state atomic temperatures: while this level of cooling has been achieved for a number of neutral atom qubit species \cite{kaufman_cooling_2012,thompson_coherence_2013,cooper_alkaline-earth_2018,norcia_microscopic_2018,jenkins_ytterbium_2022}, it is a fundamental challenge to maintain these temperatures over long sequences of gates or atom transport operations. We note that re-cooling after transport is a significant overhead in trapped ion CCD architectures \cite{bermudez_assessing_2017,pino_demonstration_2021}, and that sympathetic cooling is not straightforward in neutral atoms \cite{belyansky_nondestructive_2019}.  

\edit{}{Besides laser amplitude inhomogeneities, Doppler shifts, and decay of the Rydberg state, an experimental realization of the proposed gates will be affected by error sources not included in our analysis, including time dependent parameter fluctuations, such as laser phase noise, and uncertainties in the applied pulse shape. While our pulses are not explicitly robust to those errors, there is no indication that they are significantly more susceptible to them, compared to previous approaches such as the TO pulse. Including additional error sources in the design of robust pulses will be the subject of further investigation. Additionally, the presence of a finite blockade strength \cite{jandura_time-optimal_2022} and the effect of extra near-resonant levels such as additional hyperfine states in $^{171}$Yb \cite{ma_universal_2022} can be included in the pulse design using the optimal control techniques applied in this work.  Our work allows experimental efforts to be concentrated to errors sources against which our protocols are not robust, at the expense of error sources against which our protocols are robust.}

\begin{acknowledgments}
We are grateful to Shannon Whitlock, Hannes Pichler, Shuo Ma, Genyue Liu, Alex Burgers and Bichen Zhang for discussions. This
research has received funding from the European
Union’s Horizon 2020 research and innovation
programme under the Marie Skłodowska-Curie
grant agreement number 955479, the Horizon Europe programme HORIZON-CL4-2021-DIGITAL-EMERGING-01-30 via the project “EuRyQa -- European infrastructure for Rydberg Quantum Computing” grant agreement number 101070144 and from a state grant managed by the French National Research Agency under the Investments of the Future Program with the reference ANR-21-ESRE-0032. G. P. acknowledges support from the Institut Universitaire de
France (IUF) and the University of Strasbourg
Institute of Advanced Studies (USIAS). J.D.T acknowledges support from an ARO PECASE (W911NF-18-10215), and the ONR (N00014-20-1-2426).

\emph{Note added ---} While finalizing this work we became aware of related work in Ref.~\cite{fromonteil_protocols_2022}.
\end{acknowledgments}

\appendix

\section{Pulse Most Robust against Detuning Errors}
\label{app:pulse_most_robust_against_detunings}
In Sec.~\ref{subsec:detuning_robust_pulses} we showed that there exists no pulse $\Omega(t)$ such that the quantum state after the pulse is to first order insensitive to the detunings $\Delta_1$ and $\Delta_2$. In this appendix give a pulse which is nevertheless as robust as possible. For this we assume $\Delta_1=\Delta_2 = \Delta$, which is the case when the detuning error arises from frequency noise of the laser and not from Doppler shifts.

The fidelity of the pulse can be expanded as $F = F^{(0)} + \Delta F^{(1)} + \Delta^2 F^{(2)} + \mathcal{O}(\Delta^3)$ with $F^{(0)}$ given by Eq.~\ref{eq:fidelity} and
\begin{eqnarray}
    F^{(1)} &=& \frac{1}{2}\sqrt{F^{(0)}}\sum_{q\in\{10,01,11\}} \Re\left(e^{-i\theta_q}\braket{q|\psi_q^{(1)}}\right) \label{eq:appendix_F1} \\
    F^{(2)} &=& \frac{1}{16}\left|\sum_{q\in\{10,01,11\}} e^{-i\theta_q}\braket{q|\psi_q^{(1)}}\right|^2 \label{eq:appendix_F2}\\ \nonumber
    &+& \frac{1}{2}\sqrt{F^{(0)}}\sum_{q\in\{10,01,11\}} \Re\left(e^{-i\theta_q}\braket{q|\psi_q^{(2)}}\right)
\end{eqnarray}

We now consider a pulse with $F^{(0)} = 1$, i.e. $\ket{\psi_q^{0)}} = e^{i\theta_q}\ket{q}$. By normalization of $\ket{\psi_q}$ it must hold that $\Re(e^{-i\theta_q}\braket{q|\psi_q^{(1)}}) = \Re(\braket{\psi_q^{(0)}|\psi_q^{(1)}}) = 0$ and that $\braket{\psi_q^{(1)}|\psi_q^{(1)}} + 2\Re(e^{-i\theta_q}\braket{q|\psi_q^{(2)}}) = 0$. 
Inserting this into Eqs.~\ref{eq:appendix_F1} and \ref{eq:appendix_F2} we obtain $F^{(1)} = 0$ and
\begin{equation}
     F^{(2)} = \frac{1}{16}\left|\sum_q e^{-i\theta_q}\braket{q|\psi_q^{(1)}}\right|^2 - \frac{1}{4} \sum_q \braket{\psi_q^{(1)}|\psi_q^{(1)}}
     \label{eq:appendix_F2_2}
\end{equation}
Our goal is now to find the pulse which minimizes $-F^{(2)}$ while satisfying $F^{(0)}=1$. As a reference we calculate for the time-optimal pulse $-F^{(2)} = 3.45/\omax^2$.

We now insert the relation $e^{-i\theta_q}\braket{q|\psi_q^{(1)}} = -i\tau_q^R$ (see Eq.~\eqref{eq:first_order_detuning_error} into Eq.~\eqref{eq:appendix_F2_2} and use the fact that, because $\Delta_1=\Delta_2$, $\ket{\psi_{10}}$ and $\ket{\psi_{01}}$ are identical up to relabeling the states. We obtain $-F^{(2)} = -F_c^{(2)} - F_r^{(2)}$ with

\begin{eqnarray}
    -F_c^{(2)} &=& \frac{1}{16}\left(4(\tau_{10}^R)^2+3(\tau_{11}^R)^2-2\tau_{10}^R\tau_{11}\right) \\ 
   -F_r^{(2)} &=&\frac{1}{2}|\braket{r0|\psi_{10}^{(1)}}|^2 + \frac{1}{4}|\braket{W_+|\psi_{11}^{(1)}}|^2
\end{eqnarray}
Here, $-F_c^{(2)}$ is the conditional fidelity and measures the infidelity arising from deviations of the final state in the computational subspace while $-F_r^{(2)}$ measures the infidelity arising from population of the Rydberg state. Since $\tau_q^{R} > 0$ we see that also $-F_c^{(2)} > 0$ and thus $-F^{(2)} > 0$. 

\begin{figure}
    \centering
    \includegraphics{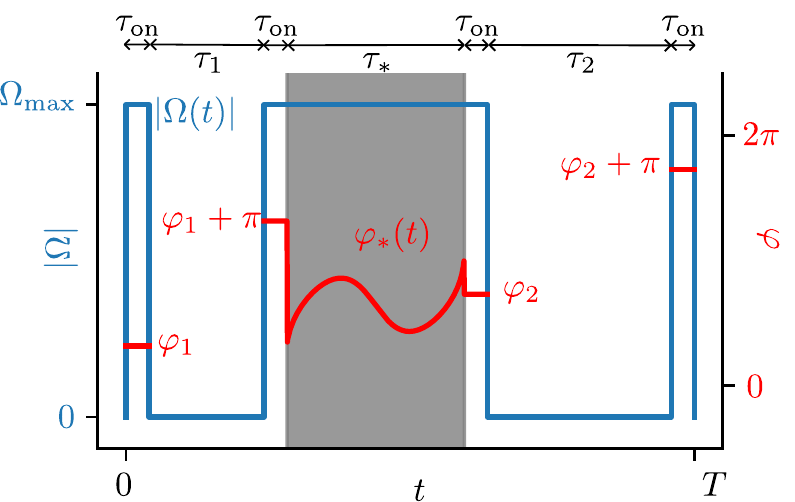}
    \caption{Schematic shape of the amplitude (blue, left vertical axis) and phase (red, right vertical axis) of the pulse achieving the highest robustness against the detuning of the laser. As explained in Appendix~\ref{app:pulse_most_robust_against_detunings} the pulse consists of 7 pieces of durations $\ton, \tau_1, \ton, \tau_*, \ton, \tau_2$ and $\ton$, respectively. The shaded area shows the pulse minimizing $-F_c^{(2)}$, while the rest of the pulse compensates the errors in the Rydberg state. }
    \label{fig:detuning_robust_pulse}
\end{figure}

A lower bound on the minimal possible $-F^{(2)}$ is obtained by minimizing $-F_c^{(2)}$ over all pulses with $F^{(0)}=1$. We do so using the GRAPE algorithm with the cost function $J = C(1-F^{(0)}) -F_c^{(2)}$ for a large $C=10^4$. The large value of $C$ ensures that the pulse minimizing $J$ will have $F^{(0)} \approx 1$, while the second term in $J$ ensures that the pulse minimizes $-F_c^{(2)}$. We find that the minimal value is $-F_c^{(2)} =2.87/\omax^2$, for a pulse $\Omega_*$ with duration $\tau_* = 7.70/\omax$. The pulse $\Omega_*$ is shown in the shaded area in Fig.~\ref{fig:detuning_robust_pulse}. Through the pulse $\Omega_*$ the value of $-F_c^{(2)}$ decreases by 17\% compared to the TO pulse.

Now we show the following: Using the pulse $\Omega_*$ as a building block we can construct a new pulse $\Omega(t)$ with $-F_r^{(2)}=0$, while still $-F_c^{(2)} =2.87/\omax^2$, the same value as for $\Omega_*$. The amplitude and phase of $\Omega(t)$ are shown schematically in Fig.~\ref{fig:detuning_robust_pulse}. The pulse is described by 5 parameters $\ton, \tau_1, \tau_2$, $\varphi_1$ and $\varphi_2$ and consists of seven parts:
\begin{equation}
    \Omega(t) = \begin{cases}
    \omax e^{i\varphi_1} & \text{ if } t \in [0, t_1] \\
    0 & \text{ if } t \in (t_1, t_2] \\
    \omax e^{i(\varphi_1+\pi)} & \text{ if } t \in (t_2, t_3] \\
    \Omega_*(t) & \text{ if } t \in (t_3, t_4] \\
     \omax e^{i\varphi_2} & \text{ if } t \in (t_4, t_5] \\
    0 & \text{ if } t \in (t_5, t_6] \\
    \omax e^{i(\varphi_2+\pi)} & \text{ if } t \in (t_6, t_7] 
    \end{cases}
\end{equation}
 with $t_1=\ton$, $t_2-t_1 = \tau_1$, $t_3-t_2 = \ton$, $t_4-t_3 = \tau_*$, $t_5-t_4 = \ton$, $t_6-t_5 = \tau_2$ and $t_7-t_6 = \ton$.   The pulse $\Omega(t)$ starts by turning on the laser for a time $\ton$ with phase $\varphi_1$, followed by an idle time of $\tau_1$ and another pulse of duration $\ton$, but this time with opposite sign of $\Omega$, i.e. with phase $\varphi_1+\pi$. After these first three parts the pulse $\Omega_*$ is applied, followed by the last three parts which have the same structure as the first three parts, but with laser phase $\varphi_2$ and idle time $\tau_2$. The pulse $\Omega(t)$ is designed such that for $\Delta=0$ it implements the same gate as $\Omega_*(t)$, because the second and the sixth part of $\Omega(t)$ have no effect and the first and third part as well as the fourth and seventh part cancel each other.
 
 In the following we consider the limit $\ton \rightarrow 0$, while keeping $\ton\tau_j$ constant (for $j=1,2$). We calculate how the pulse $\Omega(t)$ acts on the relevant computational basis states $\ket{10}, \ket{01}$ and $\ket{11}$, starting with $\ket{10}$. We start by calculating the zeroth order contribution of the state, $\ket{\psi_{10}^{(0)}}$. Outside of the short parts of the pulse with duration $\ton$ it is given by
 \begin{eqnarray}
     &&\ket{\psi_{10}^{(0)}(t)} = \\ \nonumber
     &&\begin{cases}
     \ket{10} -ie^{-i\varphi_1}\omax\ton/2 \ket{r0} & \text{ if }t\in [t_1, t_2] \\
     \ket{\psi_{10,*}^{(0)}(t-t_3)} & \text{ if }t\in [t_3, t_4] \\
     e^{i\theta_{10}}\left(\ket{10} -ie^{-i\varphi_2}\omax\ton/2 \ket{r0}\right) & \text{ if }t\in [t_5, t_6]
     \end{cases}
     \label{eq:app_a_psi10_0th_order}
 \end{eqnarray}

 where $\ket{\psi_{10,*}}$ denotes the state when executing the pulse $\Omega_*(t)$. Note that because we work in the limit $\ton \rightarrow 0$ there is no population in the Rydberg state except when the pulse $\Omega_*$ is executed, so that $\braket{10|\psi_{10}^{(1)}(t_7)} = \braket{10|\psi_{10,*}^{(1)}(\tau_*)} $ and thus $-F^{(2)}_c = 2.87/\omax^2$.
 
 To calculate the error along $\ket{r0}$,  $\braket{r0|\psi_{10}^{(1)}(t_7)}$, we note that the pulse $\Omega_*$ maps the state $\ket{r0}$ to $e^{-i\theta_{10}}\ket{r0}$ in the $\Delta=0$ case. This is due to the symmetry between $\ket{10}$ and $\ket{r0}$ in the Hamiltonian~\eqref{eq:H10}. This fact together with Eq.~\eqref{eq:app_a_psi10_0th_order} gives
 \begin{eqnarray}
     \braket{r0|\psi_{10}^{(1)}(t_7)} = &-& e^{i(-\theta_{10}-\varphi_1)}\tau_1\ton\omax/2 \\ \nonumber
     &+& \braket{r0|\psi_{10,*}^{(1)}(\tau_*)} \\ \nonumber
     &-& e^{i(\theta_{10}-\varphi_2)}\tau_2\ton\omax/2
 \end{eqnarray}
Analogously we find when starting in $\ket{11}$ that 
 \begin{eqnarray}
     \braket{W_+|\psi_{11}^{(1)}(t_7)} = &-& \sqrt{2}e^{i(-\theta_{11}-\varphi_1)}\tau_1\ton\omax/2 \\ \nonumber
     &+& \braket{W_+|\psi_{11,*}^{(1)}(\tau_*)} \\ \nonumber
     &-& \sqrt{2}e^{i(\theta_{11}-\varphi_2)}\tau_2\ton\omax/2
 \end{eqnarray}
The pulse $\Omega(t)$ thus satisfies $-F_r^{(2)}=0$ if 
\begin{equation}
     \frac{1}{2}\left(\begin{matrix}e^{-i\theta_{10}} & e^{i\theta_{10}} \\ \sqrt{2}e^{-i\theta_{11}} & \sqrt{2}e^{i\theta_{11}} \end{matrix}\right) \left(\begin{matrix}\xi_1 \\ \xi_2 \end{matrix}\right) = \left(\begin{matrix}
        \braket{r0|\psi_{10,*}^{(1)}(\tau_*)} \\
        \braket{W_+|\psi_{11,*}^{(1)}(\tau_*)}
    \end{matrix}\right)
    \label{eq:app_a_linear_system}
\end{equation}
with $\xi_j = e^{-i\phi_j}\tau_j\ton\omax$. 
Now the $\xi_j$ and thus the $\tau_j$ and $\varphi_j$ can be found by simply solving the linear system of equations~\eqref{eq:app_a_linear_system}. We find the solutions $\varphi_1 = 2.21$, $\varphi_2 = -0.05$, $\tau_1 = \tau_2 = 1.01/(\ton\omax^2)$. We numerically verified that in the limit $\ton \rightarrow 0$ the pulse $\Omega(t)$ indeed achieves $-F^{(2)} =2.87/\omax^2$.

\section{Comparing Switch and Wait Method of Doppler Shift Reversal}
\label{appendix:comparing_switch_and_wait}
In Secs.~\ref{sec:infidelities_in_a_realistic_error_model} and ~\ref{sec:conditional_infidelity_and:logical_error_rate} we saw that the switch and the wait methods give identical fidelities and logical error rates. In this appendix we discuss two aspects in which the performance of the switch and the wait differ. We start in Sec.~\ref{subsec:robustness_to_trap_inhomogeneities_and_anharmonicities} by discussing the effects of trap inhomogeneities and anharmonicities, which only affect the wait method. In Sec.~\ref{subsec:performance_without_trap_modulation} by we then show that \emph{without} the trap modulation the performance of the wait method is unaffected, while the performance of the switch method decreases significantly. We conclude with a comparison between the switch and the wait method in Sec.~\ref{subsec:comparison_between_switch_and_wait}.

\subsection{Robustness to Trap Inhomogeneities and Anharmonicity}
\label{subsec:robustness_to_trap_inhomogeneities_and_anharmonicities}

The wait method is sensitive to several non-ideal characteristics that can arise in practice. The first is an imprecise knowledge of the trap frequency, or a difference in frequency across multiple traps. This gives a contribution to the infidelity scaling as $T\sigma_\omega^2$, where $\sigma_\omega$ is the standard deviation of the trap frequency $\otr$.  We find that the induced errors are almost completely in the computational subspace, so that the contribution to the conditional infidelity is the same as to the infidelity. In order to keep a relative impact below 10\% on the infidelity at $T=50\,\mu$K, we require $\sigma_\omega/\otr< 0.06$ (0.04) for the DR (ADR) pulse. However, to maintain an erasure bias above 45 [90\% of the value shown in Fig.~\ref{fig:conditional_fidelity}(c)], the trap frequencies must be stabilized to $\sigma_\omega/\omega< 0.01$ (0.005). We note that achieving a 1\% frequency uniformity requires only 2\% intensity uniformity (assuming equal beam sizes), a number which has been experimentally demonstrated in large-scale tweezer arrays \cite{madjarov_entangling_2021}.

The wait method is also sensitive to the anharmonicity of the trap, which naturally arises from the Gaussian shape of the optical tweezer, and gives rise to a temperature-dependent trap frequency. As before, there is a similar contribution to the infidelity and conditional infidelity, which scales as $T^3/U_0^2$, where $U_0$ is the tweezer depth. Considering an optical tweezer with a $1/e^2$ intensity radius $w_0 = 500$ nm, the trap frequency $\otr = 2\pi \times 50$ kHz assumed in Secs.~\ref{sec:infidelities_in_a_realistic_error_model} and~\ref{sec:conditional_infidelity_and:logical_error_rate} is consistent with a tweezer depth of $U_0 = 127\,\mu$K, for which the anharmonicity will affect the erasure bias in Fig.~\ref{fig:conditional_fidelity}(c) significantly, reducing it to approximately $\eta_e = 27$(17) at $T=10\,\mu$K and $\eta_e =3.0$(1.8) at $T=30\,\mu K$ for the DR(ADR) pulse.  However, the conditional infidelity improves as $1/U_0^2$, allowing for rapid improvement in deeper traps. For example, setting $U_0 = 2$ mK can recover $\eta_e \approx 40$ for the ADR pulse at temperatures up to 30 $\mu$K. At fixed $w_0$, this will increase the trap frequency to $\otr = 2\pi \times 200$ kHz, which we have verified does not significantly affect the other results presented. The effect is even smaller for lighter atoms or larger $w_0$.

Note that trap inhomogenieties and anharmonicity affect the wait method regardless of the use of the trap modulation, while the switch method is unaffected by these imperfections.

\subsection{Performance without Trap Modulation}
\label{subsec:performance_without_trap_modulation}

\begin{figure}
    \centering
    \includegraphics[width=\linewidth]{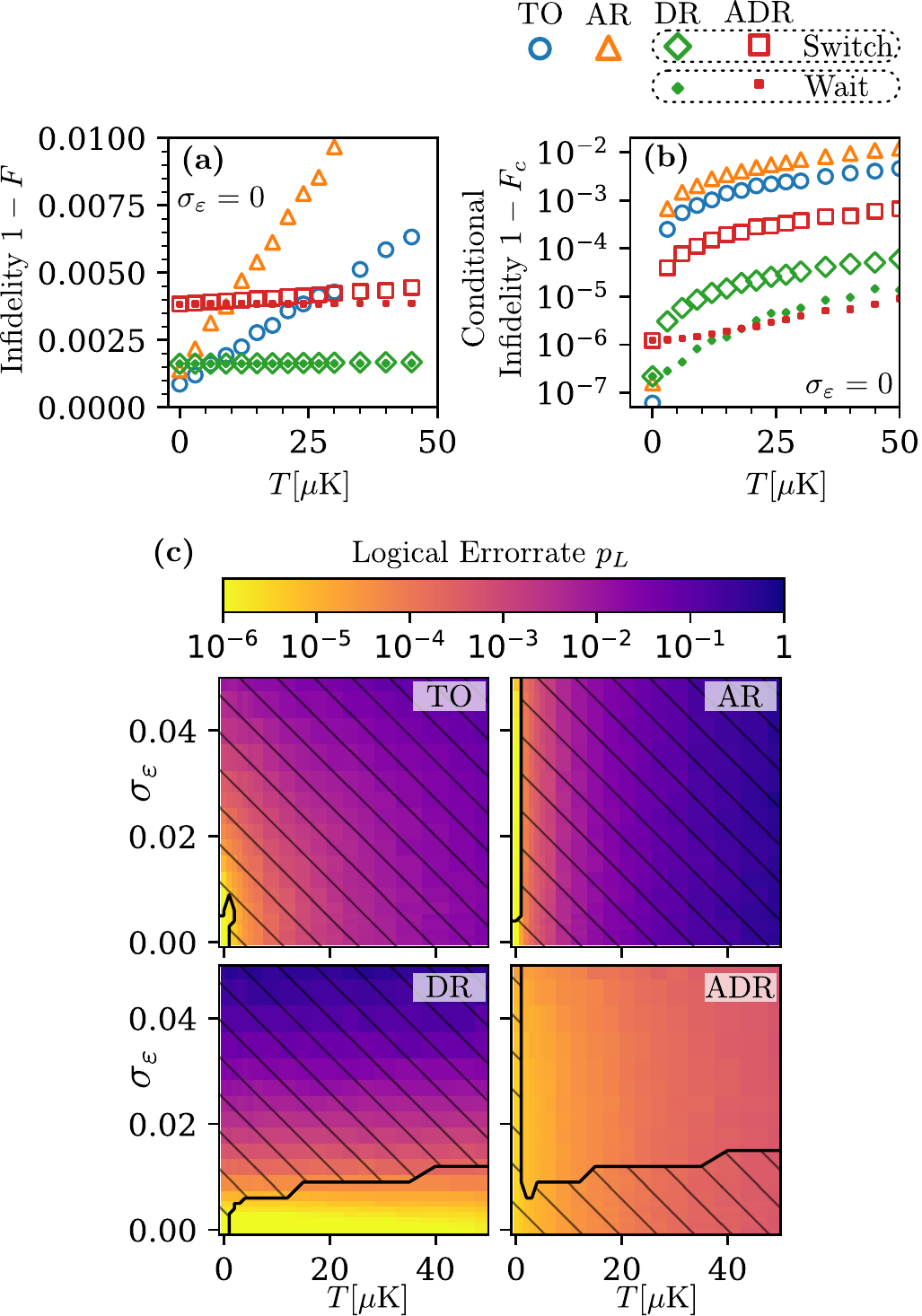}
    \caption{Performance of the switch and the wait method \emph{without} the trap modulation a) Infidelity at $\sigma_\varepsilon=0$, b) Conditional infidelity at $\sigma_\varepsilon=0$ c) The logical error rate switch method. For the ADR pulse a decrease of the logical error rate by approximately one order of magnitude is observed compared to the use of the trap modulation (Fig.~\ref{fig:conditional_fidelity}d))}
    \label{fig:performance_without_trap_modulation}
\end{figure}

In Fig.~\ref{fig:performance_without_trap_modulation}(a),(b) we show the infidelity and the conditional infidelity without the trap modulation. We observe that for the wait method the infidelity and conditional infidelity are essentially identical to the results including the trap modulation [see Figs.~\ref{fig:unconditional_fidelity}(b) and~\ref{fig:unconditional_fidelity}(c)]. In contrast, for the switch method without the trap modulation the ADR pulse has a slightly worse fidelity, and the conditional infidelity for the DR and ADR pulse increases by one to two orders of magnitude. We conclude that the switch method is significantly more sensitive to changes in the velocity during each pulse than the wait method. We attribute this to the fact that for a time dependent velocity, the wait method gives a more exact reversal of the Doppler shift than the switch method. This is because in the switch method only the Doppler shift at the end of first pulse is the negative of the Doppler shift at the beginning of the second pulse, while for the wait method the Doppler shift at each point in the first half is the negative of the Doppler shift of the same point in the second half.

In Fig.~\ref{fig:performance_without_trap_modulation}(c) the logical error rate of the switch method without the trap modulation is shown. The error rate achievable at 50 $\mu$K using the ADR pulse increases by approximately one order of magnitude to $10^{-4}$ compared to the use of the trap modulation. Again the wait method performs identically regardless of whether the trap modulation is applied (not shown).

Note that it is still favourable to apply the trap modulation for the wait method, because it mitigates differential light shifts and the anti-trapping of the Rydberg state, which are not considered above.

\subsection{Comparison between Switch and Wait Method}
\label{subsec:comparison_between_switch_and_wait}

 Combining the results from Secs.~\ref{subsec:robustness_to_trap_inhomogeneities_and_anharmonicities} and~\ref{subsec:performance_without_trap_modulation} we can summarize the advantages and disadvantages of both methods: The switch method requires a more elaborate experimental setup than the wait method, because the laser direction has to be switched. Additionally, its performance is worse than the wait method if the trap modulation is not applied. However, the switch method is more robust to trap inhomogenieties and anharmonicity. On the other hand, the wait method can be implemented with just a single laser beam, and achieves the same performance regardless of whether the trap modulation is applied (neglecting errors from differential light shifts). However, it is affected by trap inhomogenieties and anharmonicity, whose effect can be mitigated by increasing the trap frequency and depth.  

\section{Modulation of the Trapping Potential}
\label{app:modulation_of_the_trap_frequency}

\begin{figure}
    \centering
    \includegraphics[width=\linewidth]{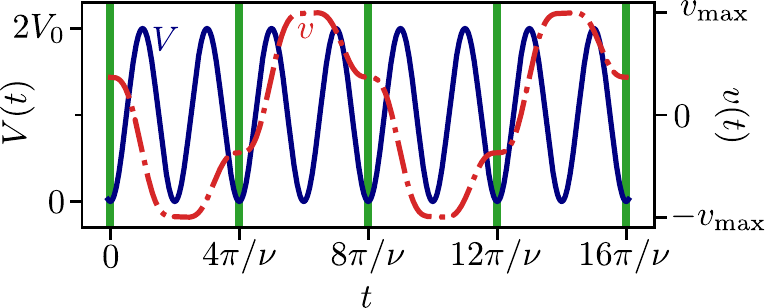}
    \caption{We propose a sinusodial modulation of the trapping potential $V$ (blue, solid line) with frequency $\nu$. The exemplary velocity of an atom moving in this potential is shown by the red, dash-dotted line. The two halves of the DR and ARD pulse are executed in adjacent time-slots marked by the green, vertical lines. During the time-slots the potential is zero, and the velocity is thus constant. The velocity switches sign between adjacent time-slots.}
    \label{fig:potential_and_velocity}
\end{figure}

The wait method proposed in Sec.~\ref{subsec:switching_the_doppler_detuning} requires a periodic motion of the atoms in the optical tweezer trap. While this can be in principle achieved simply by keeping the trapping potential constant in time, this approach induces differential light shifts between ground and Rydberg states, and can also lead to the anti-trapping of the Rydberg state. In the following we provide a method to achieve a periodic motion of the atoms through a trapping potential which is sinusoidally modulated in time. The pulse halves can then be executed at times where the trap intensity vanishes and there is no differential light shift and no anti-trapping of the Rydberg state. Additionally this method ensures that the velocity of the atoms, and thus the Doppler shift, is constant during each of the pulse halves. This is an improvement because the Doppler robust pulses are only designed to be robust against a Doppler shift which is constant during each pulse half.

We propose to modulate the potential $V$ induced by the optical tweezers trapping the atoms sinusoidally in time with frequency $\nu$, so that it is given by
\begin{equation}
    V(t,x) = V_0(1-\cos(\nu t))x^2
\end{equation}
The evolution of the atoms in the trap is thus governed by $\ddot{x} = -\frac{2V_0}{m}(1-\cos(\nu t))x$, which is a rescaled version of the Mathieu differential equation \cite{abramowitz_handbook_2013}. According to Floquet's theorem, the solutions are of the form $x(t) = e^{i\otr t}y(t)+\cc$ where $y$ is a $2\pi/\nu$-periodic function and $2\otr/\nu$ is called the Mathieu characteristic exponent. While $\otr$ can in general be complex, it has been shown that $\otr \in \mathbb{R}$ for sufficiently large $\nu$ \cite{tamir_characteristic_1962}. For example, in the limit $\nu \rightarrow \infty$ the potential $V(t)$ can be replaced by its time average and we obtain simply $\otr = \sqrt{2V_0/m}$. 

We now apply the two pulse halves that make up the DR and ADR pulse centered at times $t_1=2\pi n_1/\nu$ and $t_2=2\pi n_2/\nu$, where $n_1$ and $n_2$ are integers. In this way $V(t_1)=V(t_2)=0$, so the atoms move with a constant velocity and differential light shifts vanish. Now our goal is to find a modulation frequency $\nu$ such that a velocity reversal is achieved between $t_1$ and $t_2$. Since we require $v(t_1)=-v(t_2)$ the relation $(t_2-t_1)\otr = \pi$ has to be satisfied, so $\nu = 2(n_2-n_1)\otr$. For a given value of $V_0$ and $(n_2-n_1)$ we can now numerically find $\nu$ by first finding the Mathieu characteristic exponent $\otr$. Real solutions for $\nu$ exist for $n_2-n_1 \geq 2$. To ensure that the duration of the time slots with almost constant velocity is as long as possible we take  $n_2-n_1=2$ and find $\nu = 4.079\sqrt{2V_0/m}$. In our numerical calculations of the infidelity we took $V_0 $ such that $\otr=\nu/4$ stays at the value of $50$ kHz that we assumed before the trap modulation. The maximum potential $2V_0$ has to be roughly twice as much as the potential needed for an oscillation of the atoms at the same frequency without the trap modulation.

The potential $V(t)$ and the exemplary velocity $v$ of an atom moving in this potential are shown in Fig.~\ref{fig:potential_and_velocity}. The two pulse halves making up the Doppler robust pulse are to be applied in two adjacent time slots marked by the vertical green bars. As can be observed from the figure, the velocity is flat at these time slots, and changes sign between any two adjacent time slots. Furthermore, the differential light shift is strongly suppressed over the duration of the pulse.

\bibliography{library}

\end{document}